\documentclass[aps,prb,groupedaddress,showpacs,twocolumn]{revtex4}
\usepackage{graphicx}
\usepackage{amsmath}
\usepackage{amssymb}
\usepackage{bbm}
\usepackage[utf8]{inputenc}
\usepackage[normalem]{ulem} 

\usepackage{xspace} 
\usepackage{color}
\usepackage{footnote}

\usepackage{hyperref}

\newcommand{\beq}{\begin{equation}}
\newcommand{\eeq}{\end{equation}}
\newcommand{\beqs}{\begin{equation*}}
\newcommand{\eeqs}{\end{equation*}}

\definecolor{sigmacolor}{rgb}{0.7,0,1}

\global\long\def\eps{\varepsilon}
 \global\long\def\und#1{\underline{#1}}

\global\long\def\dag{\dagger}

\newcommand{\up}{\ensuremath{\uparrow}}
\newcommand{\dw}{\ensuremath{\downarrow}}

\global\long\def\li{{\cal L}}

\newcommand{\al}{\alpha}

\providecommand{\braket}[1]{\left<\,#1\,\right>}

\newcommand{\iim}{\Im{m}\,}

\newcommand{\vv}[1]{\boldsymbol{#1}}
\newcommand{\si}{\sigma}   

\newcommand{\nag}{{\phantom{\dag}}}
\newcommand{\se}{Sec.\@\xspace}
\newcommand{\fig}[1]{Fig.\thinspace{}\ref{#1}}

\newcommand{\eqs}[1]{Eqs.\thinspace{}(\ref{#1})}
\newcommand{\tcite}[1]{Ref.~\onlinecite{#1}}

\global\long\def\bbar#1{\overline{#1}}

\providecommand{\beqn}{\begin{eqnarray}}
\providecommand{\eeqn}{\end{eqnarray}}

\def\beqa#1\eeqa{\begin{align}#1\end{align}}

\newcommand{\Ga}[1]{\Gamma^{(#1)}}
\newcommand{\vgam}[1]{{\vv{\Gamma}}^{(#1)}}

\newcommand{\Daux}{\underline{\Delta}_\mathrm{aux}(\omega)}
\newcommand{\DRaux}{\Delta^R_\mathrm{aux}(\omega)}
\newcommand{\DKaux}{\Delta^K_\mathrm{aux}(\omega)}
\newcommand{\Dph}{\underline{\Delta}_\ph(\omega)}
\newcommand{\DKph}{\Delta^K_\ph(\omega)}
\newcommand{\DRph}{\Delta^R_\ph(\omega)}

\newcommand{\rig}{r}
\newcommand{\lef}{l}
\newcommand{\lam}{\al}

\renewcommand{\und}[1]{\uline{#1}} 

\newcommand{\figfits}{Figs.~(\ref{fits1}-\ref{fits3})\xspace}

\newcommand{\eeqref}[1]{Eq.\thinspace{}(\ref{#1})}

\definecolor{orange}{RGB}{252,77,6}
\definecolor{brown}{RGB}{200,127,50}
\definecolor{blue}{RGB}{00,000,100}
\definecolor{green1}{RGB}{00,100,00}
\definecolor{green2}{RGB}{00,150,00}
\definecolor{green3}{RGB}{00,200,00}
\definecolor{green4}{RGB}{00,250,00}
\definecolor{pink}{RGB}{255,00,127}
\definecolor{lila}{RGB}{255,00,255}

\newcommand{\ea}[1]{{ #1}}

\usepackage{ulem}
\definecolor{grey}{RGB}{100,100,100}

\newcommand{\AUX}{auxiliary reservoir\xspace}

\newcommand{\ze}{{}}
\newcommand{\aux}{\mathrm{aux}}
\newcommand{\ph}{\mathrm{ph}}
\newcommand{\iii}{\mathrm{int}}

\DeclareMathAlphabet\mathbfcal{OMS}{cmsy}{b}{n}
\newcommand{\VG}{{\mathbfcal{G}}}

\newcommand{\footnoteremember}[2]{
\footnote{#2}
\newcounter{#1}
\setcounter{#1}{\value{footnote}}
}
\newcommand{\footnoterecall}[1]{
\cite{endnote\arabic{#1}} 
}

\begin{document}

\title{
Optimized auxiliary
\ea{
representation 
}
of a non-Markovian environment \\ 
by a Lindblad equation
}

\author{Antonius Dorda}
\affiliation{Institute of Theoretical and Computational Physics, Graz University of Technology, 8010 Graz, Austria}
\author{Max E. Sorantin}
\affiliation{Institute of Theoretical and Computational Physics, Graz University of Technology, 8010 Graz, Austria}
\author{Wolfgang von der Linden}
\affiliation{Institute of Theoretical and Computational Physics, Graz University of Technology, 8010 Graz, Austria}
\author{Enrico Arrigoni}
\email[]{arrigoni@tugraz.at}
\affiliation{Institute of Theoretical and Computational Physics, Graz University of Technology, 8010 Graz, Austria}

\date{\today}

\begin{abstract}
We present a general scheme to map correlated nonequilibrium quantum
impurity problems onto an auxiliary open quantum system of small
size. The infinite fermionic reservoirs of the original system are
thereby replaced by a small number $N_B$ of noninteracting auxiliary
bath sites whose dynamics is described by a Lindblad equation.  Due to
the presence of the intermediate bath sites, the overall dynamics
acting on the impurity site is non-Markovian.

With the help of an optimization scheme for the auxiliary Lindblad
parameters, an accurate mapping is achieved, which becomes
exponentially exact upon increasing $N_B$.  The basic idea for this
scheme was presented previously in the context of nonequilibrium
dynamical mean field theory.  In successive works on improved manybody
solution strategies for the auxiliary Lindblad equation, such as
Lanczos exact diagonalization or matrix product states, we applied the
approach to study the nonequilibrium Kondo regime.

 In the present paper, we address in detail the mapping procedure
itself, rather than the many-body solution.  In particular, we
investigate the effects of the geometry of the auxiliary system on the
accuracy of the mapping for given $N_B$.  Specifically, we present a
detailed convergence study for five different geometries which,
besides being of practical utility, reveals important insights into
the underlying mechanisms of the mapping.  For setups with onsite or
nearest-neighbor Lindblad parameters we find that a representation
adopting two separate bath chains is by far more accurate with respect
to other choices based on a single chain or a commonly used star
geometry.  A significant improvement is obtained by allowing for
long-ranged and complex Lindblad parameters.  These results can be of
great value when studying Lindblad-type approaches to correlated
systems.

\end{abstract}


\maketitle

\newcommand{\flab}{{f}} 

\section{Introduction}\label{intro}

Strongly correlated systems out of equilibrium have recently attracted
considerable interest due to 
 progress in several experimental fields, such as ultrafast pump-probe spectroscopy~\cite{iw.on.03,ca.de.04}, ultracold quantum gases~\cite{ra.sa.97,ja.br.98,gr.ma.02,tr.ch.08,sc.ha.12}, solid-state nanotechnology~\cite{bo.gr.05,go.go.98, kr.sh.12}.
These advances have also prompted the interest in related theoretical questions 
concerning thermalisation~\cite{caza.06,ca.ca.07,ri.du.08}, dissipation and decoherence~\cite{le.ch.87}, and nonequilibrium quantum phase transitions~\cite{mi.ta.06}.
An interesting aspect is the interplay between correlation and
dissipation in systems in which the latter is not included 
phenomenologically 
but is 
part of the microscopic model.
The challenge lies in the fact that the Hilbert space for correlated fermionic systems
increases exponentially with  system size.
For a finite system, on the other hand,
the  spectrum remains discrete and
dissipation does not occur.
When considering purely fermionic correlated systems, dissipation 
is usually modeled  by  infinite reservoirs of noninteracting fermions.
These reservoirs are in contact with a correlated central region of interest.
A paradigmatic example of such a system is the single-site Kondo
or Anderson impurity 
model~\cite{hews}.
If there is just one  reservoir with a single chemical potential $\mu$ and
temperature $T$, then the whole system (typically) reaches thermodynamic equilibrium.
Alternatively,
one can consider a nonequilibrium situation in which several reservoirs with
different $\mu$ and $T$ are in contact with the central region. 
Since the reservoirs are infinite they act as dissipators and
the system  in most cases reaches 
 a nonequilibrium 
steady
state in which a particle and/or heat current flows across the
central region. \footnote{A notable exception is when a bound state is present, i.e. a state with energy outside of the continuum of the reservoir. In this case, there is no unique  steady state.}

There are  
\ea{
several 
}
 approaches to treat such systems numerically.
Some of them start out from the situation in which 
 the central region and the reservoirs are decoupled which allows the  individual systems to be  treated exactly. There are different 
\ea{
schemes 
}
 to include the missing coupling  between the reservoirs and the central region.
First of all, one could carry out a perturbative expansion in terms of the
reservoir-central region coupling. 
Low energy properties are better addressed within 
 a renormalisation-group treatment of the perturbation (see,
e.g. Ref.~\onlinecite{scho.09}). 
Alternatively, one can try and compute the self-energy 
\ea{
(most  
}
 nonequilibrium quantities of interest follow from Dyson's equation) 
for the correlated sites based on finite clusters 
consisting of the central region plus a small number $N_{r}$ of reservoir sites.
 This is done in nonequilibrium cluster 
perturbation
theory~\cite{ba.po.11,nu.do.15},  
whose accuracy increases with increasing $N_{r}$. A generalization of this idea  is the nonequilibrium variational cluster 
approach,~\cite{kn.li.11,nu.ar.12,ho.ec.13}, where 
\ea{
single-particle 
}
 parameters of the model are optimized self-consistently, which allows for the adjustment of the self-energy to the nonequilibrium situation.

In a different type of approach one tries
 to ``eliminate'' the degrees of
freedom of the reservoir and take into account its effects on the
dynamics of the interacting central region. Formally, this can be expressed
in terms of a functional integral whereby the part of the action
describing the reservoirs, which is quadratic, is integrated out and
one obtains an effective action restricted to the central region only,
whereby the effects of the reservoir introduce couplings with
retardation effects. These physically describe processes in which
particles jump from the central region to the reservoir and then come
back after a certain delay. 
This retarded action 
can
 be treated, e.g.,  via continuous time Monte Carlo
approaches~\cite{ru.sa.05,we.co.06}, which, however are plagued by the minus sign problem.
Due to retardation, exact diagonalization approaches are not appropriate.
There are several other ways to achieve this elimination of 
the reservoir degrees of freedom. 
Renormalisation-group approaches are certainly convenient
whenever one is interested in the low-energy sector.~\cite{ja.me.07,ge.pr.07} 
The numerical
renormalisation group has proven extremely powerful for quantum
impurity models.~\cite{hews}  

\subsubsection{Markovian approximations and beyond}
Another approach consists in treating the
coupling to the reservoir within the Born-Markov approximation. In this way, the
effect of the reservoir is to introduce nonunitary dynamics in the
time dependence of the reduced density operator $\rho_\flab$ of the central region
 leading to the Lindblad equation~\cite{br.pe}, which is a
linear, time-local equation for $\rho_\flab$ preserving its hermiticity, trace, and
positivity.
One important precondition for the validity of this mapping, however, is
the Markovian  assumption
that the  decay of correlations in the reservoir
is
 much faster than typical time scales of the central region.~\cite{br.pe} 
As pointed out, e.g. in  
 Refs.~\onlinecite{cohentann,br.pe} the approximations
leading to the Markovian Lindblad master equation
are justified provided
the typical energy scale $\Omega$ of the reservoir is much larger than
the reservoir-central region coupling.
However, for a fermionic system, 
$\Omega$ can be estimated as $\min(W,\max(|\mu-\eps|,T)$, where $W$
is the reservoir's bandwidth, and $\eps$ is a typical single-particle
energy of the central region. Therefore, even in the wide-band limit $W\to\infty$,
the validity of the Markov approximation is limited either to
high temperatures or to chemical potentials far away from the
characteristic energies of the central region. 
As a matter of facts,
the effect of a
noninteracting reservoir with $W,|\mu|\to\infty$
(or $T\to\infty$ with finite $\mu/T$) can be exactly written in terms of a Lindblad
equation. This can be easily deduced from the ``singular coupling'' derivation
of the Lindblad equation~\cite{br.pe}.
This is valid independently of the strength of the coupling 
between central region and reservoir. 
A nontrivial situation is obtained by introducing different reservoirs with different particle densities.
The pleasant aspect of this limit is
that the Lindblad parameters depend on the properties of the reservoir
and of its coupling with the central region  only, but not on the ones of the central region.

This is in contrast to the more standard 
weak-coupling Born-Markov
 version in
which the Lindblad couplings (see, e.g.~\cite{br.pe,scha}) depend on
the central region's properties. 
\ea{
To illustrate this, consider 
}
 a central region consisting of a single site with energy $\eps_f$,
i.e. with  Hamiltonian 
\beq
H_\flab= \eps_f \ f^\dag f
\eeq
 (omitting spin)
and reduced density matrix 
\ea{
$\rho_f$.  The 
}
part of the Lindblad operator $\li_b$
 describing the coupling to a noninteracting reservoir 
 is given by 
\beqa
\label{li1}
\li_b \ \rho_\flab &=  \Gamma_1 \left(2 f \rho_\flab f^\dag  -\{f^\dag f,\rho_\flab \}\right)
\nonumber \\ &
+
\Gamma_2 \left(
2 f^\dag \rho_\flab f  -\{f f^\dag,\rho_\flab \}
\right)
\eeqa
with 
\beq
\label{li1g}
\Gamma_1=\Gamma(1-f_F(\eps_f)) \quad\quad
\Gamma_2=\Gamma\ f_F(\eps_f)  \;.
\eeq
Here, $\Gamma$ is proportional to the reservoir's density of states at the energy $\eps_f$, and
 $f_F$ is the Fermi function which obviously contains the
information 
on
 the chemical potential and temperature of the reservoir
but also on the 
onsite energy $\eps_f$ in the central region.
This could be
unsatisfactory since one would like to describe the effect of the
reservoir in a form which is independent of the properties of the
central region, 
 especially
 when the latter consists of
many coupled sites.

One possible way to 
eliminate the dependence of the Lindblad couplings on the parameters of the central region
  is to use
 an intermediate auxiliary
buffer zone (mesoreservoir) between the Lindblad couplings and the central region (see, e.g. \cite{dz.ko.11sa,dz.ko.11,aj.ba.12})
The buffer zone  consists of isolated discrete sites (levels)
each one coupled to a Markovian environment described by Lindblad operators
with the same $T$ and $\mu$ 
as given in \eeqref{li1},\eeqref{li1g}.
If the buffer zone is sufficiently large, i.e. if its
 levels are dense enough,
then one can show that 
the buffer zone including Lindblad operators yields an accurate
representation of the reservoir,
 which becomes exact in the limit of an infinite number of levels.
 Importantly, the parameters of this 
buffer zone 
 do not depend on the central region's properties.
The disadvantage of this approach is that one needs  quite a large
number of buffer levels, especially at low temperatures where the
Fermi function is sharp. Consequently, the many-body Hilbert space
is too large and the treatment of a correlated problem becomes prohibitive.

\subsubsection{This work}

In this paper, we show that the accuracy of the buffer-zone idea
can be improved significantly
even with a moderate number of auxiliary 
buffer levels (sites) 
 by allowing for more 
\ea{
general 
}
 Lindblad couplings, which are adjusted to optimize the representation of the physical reservoirs independently of the parameters of the central region. 
In particular, we show that allowing for long-ranged, 
\ea{
and even complex 
}
Lindblad terms (see below) dramatically improves the accuracy of the reservoir's description for a fixed  number of  auxiliary sites. 
In the case of a single-impurity model, already a small number of sites ($4$ to $6$) is 
\ea{
enough in order 
}
to reach a very good accuracy~\cite{do.nu.14,ti.do.15,do.ti.16} 
sufficient
 to resolve the splitting of the Kondo peak at finite bias. 
This is crucial, since the Hilbert space of such a small system can  still be treated by Krylov-space methods~\footnoteremember{kdens}{Consider that within 
the many-body Lindblad equation we have to deal with the space of density matrices not of state vectors, therefore the number of sites that can be dealt with is smaller.
}
Larger systems can be tackled by matrix-product-state approaches for open quantum systems~\cite{ve.ga.04,zw.vi.04,pr.zn.09,do.ga.15}. 
Due to its rapid convergence  this scheme can be used as an accurate  nonequilibrium solver for  
 correlated impurity problems, which even in equilibrium becomes competitive with other established approaches.
\ea{
Here  we discuss in particular 
}
several way to optimally
 represent a  physical (``ph'') reservoir by means of an auxiliary (``aux'') one consisting  of a small number of noninteracting fermionic sites and arbitrary Lindblad terms.
Presenting results of  the application to  interacting systems is not the main goal 
of this work, so there will be only a short discussion in Sec.~\ref{inte}, and we refer to previous publications~\cite{do.nu.14,do.ga.15,ti.do.15} for details. 
Here, we are interested in a systematic analysis of the performance of different geometries of the \AUX, see Fig.~\ref{fig:setups_sketch}, including a scaling analysis of the accuracy as a function of the number of bath sites $N_B$ and  a discussion of the importance of long-ranged Lindblad terms.

This paper is organized as follows: in Sec.~\ref{model}, we introduce the 
 models we are interested in and define the basic notation. In Sec.~\ref{mapp}, we 
\ea{
illustrate 
}
the most important aspect of this work, namely the mapping of the physical Hamiltonian problem onto an auxiliary open quantum system described by a Lindblad equation. In Sec.~\ref{aux}, we present the expressions for the non-interacting Green's function of the auxiliary system, and in Sec.~\ref{fit} we illustrate the fit procedure. In Sec.~\ref{inte}, we briefly discuss the relation with the interacting case. 
In Sec.~\ref{sec:results},  we present in detail the convergence of the fit as a function of 
 $N_B$  for the different geometries presented in Sec.~\ref{sec:setups} and for different temperatures, and discuss 
\ea{
 the 
}
 advantages and disadvantages of these setups.
Finally, in Sec.~\ref{summ} we summarize our results and discuss possible improvements and open issues.
In three appendices we present technical details of the minimization procedure (Sec.~\ref{app:PT}), show the explicit form of the matrices for the different geometries  
(Sec.~\ref{mat}), 
and discuss certain redundancies of the auxiliary system (Sec.~\ref{rstar}).

\section{Model and Method}
\subsection{Model}
\label{model}

We begin with a general discussion, which we eventually apply to the single-site Anderson impurity model. In the general case the central region may represent a small cluster or molecule.
The Hamiltonian of the physical system at study is written as
\beq
\label{hfull}
H = \sum_{\al} \left( H_{\al} + H_{\al \flab} \right) + H_\flab 
\eeq
where 
$H_\flab$ is
the  Hamiltonian 
of the central region 
describing a small cluster of
interacting fermions,
$H_{\al}$ is the  Hamiltonian 
of the reservoir $\al$
describing an
infinite lattice of noninteracting particles, and $H_{\al \flab}$ is the
coupling between central region and reservoirs.
\beq
\label{hclab}
H_\flab = H_{0\flab} + H_U
\eeq
consists of a noninteracting part
\beq
H_{0\flab} = \sum_{ij} h_{ij} f^\dag_i f_j
\eeq
and an interaction term $H_U$. 
The 
fermions in the reservoirs
 can be 
described by
\beq
\label{hal}
H_\al = \sum_{p,p'} \eps_{\al p ,p'} d^\dag_{\al p} d_{\al p'}
\eeq
in usual notation. For simplicity, spin indices are 
not explicitly mentioned
 here.
Quite generally, a suitable single particle basis ``star representation'' can be chosen such that 
$\eps_{\al p ,p'} \sim \delta_{p,p'}$.
$H_{\al \flab}$ is 
taken to be quadratic in the fermion
operators:
\beq
H_{\al \flab} = \sum_{p,i} v_{\al p i} d^\dag_{\al p} f_i +h.c. \;,
\eeq
and $d_i$ ($f_i$) are the fermionic destruction operators on the reservoir's (central region's) sites $i$.

We are interested in a steady state situation, although the present
approach can be easily extended to include time dependence, 
especially if 
this comes from a change of the central region's parameters only.
In the steady state
we can Fourier transform with respect to the time variables  so that the Green's functions depend on a real frequency $\omega$ only, which here is kept implicit.
We assume 
that initially  the hybridization $H_{\al \flab}$ is zero and
 the reservoirs  are separately
 in equilibrium with chemical
potentials $\mu_\al$ and temperatures $T_\al$.
Then  the $H_{\al \flab}$ are switched on and 
after a certain time a steady
state is reached.
We use 
 the non-equilibrium (Keldysh) formalism~\cite{kad.baym,schw.61,keld.65,ha.ja,ra.sm.86} 
whereby the Green's function 
 can be represented as a $2\times 2$ 
block matrix
\begin{align}
\label{gund}
  \und{G} &= \begin{pmatrix} G^R & G^K \\ 0 & G^A \end{pmatrix}\,\mbox{,}
\end{align}
where the retarded $G^R$, advanced $G^A$, and Keldysh 
$G^K$ components are matrices in the site indices $(i,j)$ of the central region.
We will adopt 
the above underline notation   in order to denote this $2\times 2$ structure.
We use lowercase $\und g$ ($\und g_{\al}$) to denote Green's function of the decoupled 
central region (reservoir $\al$), while uppercase $\und G$ represent the full noninteracting Green's function of the central region.
For simplicity we omit the subscript ${}_0$,  since  in this paper
we deal mainly with noninteracting Green's functions
anyway. We use the subscript ${}_\iii$ for interacting ones.
 $\und G_\ze$ is easily obtained from the Dyson equation as
\beq
\label{G0}
\und G_\ze = \left(\und g^{-1} - \und \Delta \right)^{-1} \;,
\eeq
where
\beq
\label{delij}
\und\Delta_{ij} = \sum_{\al,p,p'} v_{\al p i} \ v_{\al p' j} \ \und g_{\al p,p'}
\eeq
is the
 reservoir hybridization function 
(commonly called bath hybridization function)
in the Keldysh representation.
The retarded Green's functions
 $g^R_{\al}$ for reservoirs with non-interacting fermions in equilibrium can be determined 
easily by standard tools, and the Keldysh components 
$g^K_{\al}$ can be obtained from the retarded ones by exploiting the fluctuation-dissipation theorem:
\beq
\label{gkal}
g^K_{\al}(\omega) =  \left(g^R_{\al}(\omega) -   g^R_{\al}(\omega)^\dag \right)  \ s_\al(\omega)
\eeq
which is valid since the 
uncoupled reservoirs are in equilibrium. Here,
\beq
s_\al(\omega) = 1-2 \ f_F(\omega,\mu_\al,T_\al)
\eeq
and $f_F(\omega,\mu_\al,T_\al)$ is the Fermi function at chemical potential $\mu_\al$ and temperature $T_\al$.

From now on, for simplicity of presentation, we restrict to the  Anderson impurity model (SIAM)  in which 
the central region, described by \eeqref{hclab}, consists of a single 
 site, i.e. there is only one value for the index $i$, which we drop, and
\beq
H_U =  U n_{f\up} n_{f\dw} 
\quad \quad n_{f\si} 
= f^\dag_\si f_\si \;.
\eeq

The idea we are going to present in Sec.~\ref{mapp} can be immediately extended to the case of a central region consisting of many sites in which each site is connected to separate reservoirs. In the most direct fashion, this can be done with exactly the same approach as formulated here for the SIAM by just mapping each reservoir independently onto auxiliary Lindblad bath sites. 
An interesting application is
for example, 
the case of an interacting chain coupled on both sides to reservoirs with different chemical potentials~\cite{be.ca.09}.
Also the extension to the case of  arbitrary (quadratic) couplings with the reservoirs that intermix the central region sites, relevant, e.g., for cluster-DMFT, is conceptually straightforward, although more complex.

\subsection{Mapping onto an auxiliary master equation}
\label{mapp}
A crucial point 
in the following considerartions
 is the fact that, 
even in the interacting case,
the influence of the reservoirs upon the central region is completely determined by
the 
bath hybridization function $\und \Delta(\omega)$ only.
In other words, {\em any interacting 
 correlation function of the central region  solely depends on
the central region Hamiltonian $H_\flab$ and on $\und \Delta$}.
This result is well known, at least in equilibrium, and can be easily
proven, for example diagrammatically.(see footnote \footnoteremember{justdel}{
Since there are no interactions in the reservoir, external indices of any bare interaction vertex belong to the central region only.
Moreover, an interacting   correlation function of the central region 
 consists, by definition, 
of diagrams whose 
external lines
belong to the central region. Consequently, all diagrams consist only of vertices (determined by $H_U$) and of 
propagator lines whose endpoints  belong to the central region only, i.e. 
they correspond to
noninteracting Green's functions 
$\und G$
 of the
central region \eeqref{G0}. 
 Therefore, all relevant diagrams only depend on $H_U$ and the noninteracting
$\und G = (\und g^{-1} - \und \Delta )^{-1} $.
})
The argument holds independently on whether one works with equilibrium or nonequilibrium Green's functions. Moreover, it crucially depends on the fact that the reservoir is noninteracting.

This can be exploited to choose different representations for the reservoir depending on convenience. 
{\em In equilibrium}, especially in connection with numerical renormalisation group (NRG), one uses either the diagonal (``star'') representation in which the $\eps_{\al p,p'}\sim \delta_{p,p'}$ are diagonal, as in \eeqref{hal}, or the ``chain'' representation in which they describe a nearest-neighbor chain (see, e.g. Ref.~\onlinecite{bu.co.08}).
While for a continuous density of states one needs, in principle,  an infinite number of sites for the reservoirs, 
one can approximate the physical~\footnoteremember{phaux}{We shall use suffixes 
${}_\ph$ and ${}_\aux$ to distinguish between the physical bath hybridisation function \eeqref{delij} and the ones produced by the \AUX.
Whenever necessary to avoid ambiguities, these suffixes will be used also for other quantities.
} 
 $\und \Delta_\ph$ \eeqref{delij}
by an {\it auxiliary}  $\und \Delta_\aux$ 
corresponding to  a
bath with a  finite number of sites and
 optimize their parameters $\eps_{\al p p'}$ and $v_{\al p}$ via a best fit. 
Notice that {\em for this Hamiltonian representation}
the space of parameters is redundant, so that one can restrict, for example to diagonal
$\eps_{\al p p'}\sim \delta_{p,p'}$ and real $v_{\al p j}$. This is the ``star'' representation mentioned above.
The ``chain'' representation is given by a single nonvanishing $v_{\al p j}$ and local or nearest-neighbor $\eps_{\al p p'}$ 
\ea{
and is obtained from the star representation via a unitary transformation.  
}
 Therefore, for $N_B$ sites of the auxiliary system
 one has $2 N_B$ parameters available for the fit.
This approach is used, for example, for exact-diagonalisation-based Dynamical-Mean-Field Theory (ED-DMFT) ~\cite{ca.kr.94,ge.ko.96}. Here, the 
 parameters are optimized by fitting
the bath hybridization function in Matsubara space.
The auxiliary system of bath sites plus impurity is then solved
by Lanczos exact diagonalisation (ED).~\cite{lanc.50}.

Clearly, the same fit strategy is inconvenient {\em out of equilibrium} for several reasons. First of all, the auxiliary system cannot dissipate, since it is finite, and a steady state  cannot be reached.
In Refs.~\cite{ar.kn.13,do.nu.14} we have suggested a different approach (Auxiliary Master Equation Approach, AMEA), which adopts an \AUX, consisting of  a certain number $N_B$ of bath sites which are
additionally coupled to Markovian environments described by a Lindblad equation
\beq
 \label{eq:Leq}
 \frac{d}{dt}\rho = \li \ \rho  = \left( \li_H + \li_D\right) \rho  \,.
\eeq
Here, the  Hamiltonian for the auxiliary system 
is given by (we reintroduce spin)
\beq
 \label{eq:H_aux}
   H_{\aux} = \sum_{ij\si} E_{ij} c_{i\si}^\dagger c^\nag_{j\si} + U n_{f\up} n_{f\dw} \,,
\eeq
and enters the unitary part of the Lindblad operator  
\begin{equation}
 \label{eq:L_H}
 \li_H\ \rho = -i[H_{\aux},\rho] \,.
\end{equation}
The dissipator  $\li_D$ describes the coupling of the auxiliary sites to the Markovian environment and is given by
\begin{align}
   \li_D \rho &= 2 \sum_{ij\si} \Ga1_{ij} \left( c^\nag_{j\si}\rho c_{i\si}^\dagger - \frac{1}{2} \left\{\rho, c_{i\si}^\dagger c^\nag_{j\si} \right\} \right)  \nonumber \\
 \label{eq:L_D}
   &\hspace{1em}+ 2 \sum_{ij\si} \Ga2_{ij} \left( c_{i\si}^\dagger\rho c^\nag_{j\si}  - \frac{1}{2} \left\{\rho, c^\nag_{j\si}c_{i\si}^\dagger \right\} \right)   \,.
\end{align}
The indices $i,j$ in \eeqref{eq:L_H}, \eeqref{eq:L_D} run over the impurity $i=f$ 
(we identify $c_{f\si}=f_\si$) and over the $N_B$ bath sites.~\footnoteremember{lexact}{
It can be easily shown that the dissipative form \eeqref{eq:L_D} 
exactly corresponds
to the coupling of the auxiliary baths to a small number of {\it noninteracting reservoirs} with constant density of states and infinite chemical potentials and/or temperatures. 
This can be easily deduced from the ``singular coupling'' derivation
of the Lindblad equation~\cite{br.pe}.
}
Similarly  to the case of the ED-DMFT impurity solver mentioned above, 
the idea is to optimize the parameters 
of the \AUX 
in order to achieve a best 
  fit to the physical 
bath hybridization function \eeqref{delij}, i.e., for a given $N_B$,   $\Daux$  should  be as close as possible to  $\Dph$:
\begin{equation}
\Daux \approx \Dph \,,
\label{dphdaux}
\end{equation}
 As for the ED-DMFT case, one can choose a single-particle basis for the 
auxiliary bath 
such that the matrix $\vv E$ is sparse,~\footnoteremember{bold}{We use boldface to denote  matrices in the indices $i,j$ of the auxiliary system. This should not be confused with \underline{\dots}, which denotes matrices in Keldysh space}
 i.e. it has
 a ``star'' or a ``chain'' form, 
\ea{
and is real valued.  
}
However, there is no reason why 
the matrices $\vgam1$ and $\vgam2$ should be sparse 
\ea{
and real  
}
in the same basis as well, and, in fact, as discussed below, for an ED treatment of the Lindblad problem it is convenient to allow for a general form  
in order to  optimize the  fit.
This larger number of parameters 
allows one to fulfill \eeqref{dphdaux} to a very good approximation. The introduction of dissipators \eqref{eq:L_D}
additionally allows to carry out 
 the fit directly for real $\omega$, see \se\ref{fit} below, 
 since $\Daux$ is a continuous function. This makes this approach competitive with ED-DMFT for the equilibrium case as well.

Notice that \eeqref{eq:L_D} is not the most general form of the dissipator, and one could think of including Lindblad terms that contain four or more fermionic operators, or also anomalous and spin-flip terms. This would increase the number of parameters available for the fit. However, the latter would violate conserved quantities and the former would describe an {\em interacting bath}, so that the argument of 
Sec.~\ref{mapp} (footnote~\footnoterecall{justdel})
does not apply.
As a matter of fact, 
 the exact equivalence to a {\em noninteracting bath}~\footnoterecall{lexact} only holds 
for a 
quadratic form of the Lindblad operator as in \eeqref{eq:L_D}.

Once the optimal values of the matrices $\vv{E}$, $\vgam1$ and $\vgam2$ for a given  physical model are 
determined for the non-interacting system, 
\ea{
one could  
}
 solve for the 
 dynamics of the correlated auxiliary system
 defined by
\eeqref{eq:Leq}, which
amounts to  a linear equation 
for the reduced many-body density 
matrix.
 If the number of sites of this system is small, one can solve exactly for the steady state and the dynamics of this  interacting system by 
methods such as Lanczos exact diagonalization or 
Matrix Product States (MPS)~\cite{ve.ga.04,zw.vi.04,do.ga.15,pr.zn.09}.

\subsection{Computation of the Auxiliary bath hybridization function}
\label{aux}
In order to carry out the fit \eeqref{dphdaux}, 
we need to compute the \AUX hybridization function $\Daux$ for many values of the 
bath and Lindblad parameters.
This can be done in an efficient manner since only noninteracting Green's functions are needed, see also \eeqref{G0} and the discussion above. Computing the  single-particle Green's function
matrix $\und \VG$
 of \eeqref{eq:Leq} 
amounts to solving a noninteracting fermion problem, which scales polynomially with respect to the single-particle Hilbert space $N_B+1$. 
A method to deal with quadratic fermions with linear dissipation based on a so-called ``third quantisation'' has been introduced in \tcite{pros.08}. We adopt the approach of \tcite{dz.ko.11} in which the authors recast  an open quantum problem like \eeqref{eq:Leq} into a standard operator problem in an augmented fermion Fock space with twice as
many sites and with a non-Hermitian Hamiltonian.~\cite{dz.ko.11,schm.78,ha.mu.08} This so-called super-fermionic representation is convenient for our purposes, not only to solve for the noninteracting Green's functions but also to treat the many-body problem in an analogous framework to Hermitian problems. An analytic expression for the the noninteracting steady-state retarded and Keldysh auxiliary Green's functions was derived in \tcite{do.nu.14}. 
An alternative derivation, which does not rely on super-fermions is given in  
\tcite{sc.go.16u}.
For the retarded component 
 we get~\footnoterecall{bold}
\begin{align}
\label{gro}
\VG^R_\ze(\omega)  &= \left(\omega -\vv E+i(\vgam2+\vgam1) \right)^{-1}\,\mbox{,}
\end{align}
and the Keldysh component of the inverse Green's function reads
\begin{align}
\label{gko}
\left({\und \VG^{-1}}\right)^K 
&
= 
-2i  \left(\vgam2-\vgam1 \right) \,\mbox{,}
\end{align}
yielding the Keldysh Green's function
\beqa
\VG_{\ze}^K =&
- \VG_\ze^R \left(\vv{\und \VG^{-1}_\ze}\right)^K \VG_\ze^A 
\nonumber 
\\ 
\label{gkeff}
& =
 2i \VG_\ze^R \left( \vgam2-\vgam1 \right) \VG_\ze^A \,.
\eeqa
The $ff$ component of $\und{\VG}$ is the auxiliary impurity Green's function 
\beq
{\und G}_\aux = \left(\und{\VG}\right)_{ff} \;.
\eeq
From this one can determine  
the retarded  component of $\Daux$ 
\beq
\DRaux = 1/g^R -  1/G_\aux^R \,.
\label{eq:DRaux}
\eeq
For the Keldysh component, one has to carry out two inversions of Keldysh matrices 
(see, e.g. \tcite{ha.ja})
yielding
\beq
\DKaux = -\left({\und G}_\aux^{-1}\right)^K = 1/|G_{\aux}^R|^2  \ G_{\aux}^K \,,
\label{eq:DKaux}
\eeq
where the contribution from $g^K$ is infinitesimal.

\subsection{Fit procedure}
\label{fit}
From the equations above we can efficiently compute $\Daux$ for a given set of parameters of the \AUX.  
The numerical effort for a single evaluation is low and scales only at most as $\mathcal{O}(N_B^3)$. 
 We introduce a  vector of parameters $\vv{x}$ which yields a unique set of matrices $\vv{E}$, $\vgam1$ and $\vgam2$ within a chosen subset (see, e.g. Fig:~\ref{fig:setups_sketch} and App.~\ref{mat}), quantify the deviation from \eeqref{dphdaux} through a cost function
\begin{align}
\chi(\vv{x})^2 &= \frac{1}{\chi_0^2} \int\limits_{-\omega_c}^{\omega_c}\left\lVert \und\Delta_\ph - \und\Delta_\aux \right\rVert W(\omega) d\omega \,, \nonumber \\
\left\lVert \und\Delta_\ph - \und\Delta_\aux \right\rVert &= \hspace{-0.9em} \sum\limits_{\xi\in\{R,K\}}\hspace{-0.9em} \iim\{ \Delta^\xi_\ph(\omega) - \Delta^\xi_\aux(\omega;\vv{x}) \}^2  \,,
\label{eq:costfunc}
\end{align}
and minimize $\chi(\vv{x})$ with respect to $\vv x$. The normalization $\chi_0$ is hereby chosen such that $\chi(\vv{x})=1$ when $\Daux\equiv0$.
 It is important to note that both, the retarded and the Keldysh component must be fitted.
Due to Kramers-Kronig relations, the real part of 
$\DRph$
is fully determined by its imaginary part, provided the asymptotic behavior is fixed. Therefore, we can restrict to fit its  imaginary part, while $\DKph$ is purely imaginary.
 Furthermore, in \eeqref{eq:costfunc} we introduced a cut-off frequency
 $\omega_c$  and a   weighting function $W(\omega)$. In this work we take  $W(\omega)=1$ and $\omega_c = 1.5\,D$, with $D$ the half-bandwidth of $\Dph$. 
Different forms of $W(\omega)$ can be used in order to increase for instance the 
accuracy of the fit near the chemical potentials.
The minimization of \eeqref{eq:costfunc} constitutes a multi-dimensional optimization problem and appropriate numerical methods for it are discussed in \se\ref{app:PT}.

As asymptotic limit we require here $\Daux\to0$ for $\omega\to\pm\infty$, which is obtained when setting 
\ea{
$\Ga{1/2}_{ff}=0$.  Semipositivity further requires
$\Ga{1/2}_{if}=\Ga{1/2}_{fi}=0$.
}
For simplicity, we restrict here to the particle-hole symmetric case.
This reduces the number of free parameters in $\vv{E}$, $\vgam1$ and $\vgam2$. For the case that the impurity site $f$ is located in the center and that one has an even number of bath sites $N_B$, particle-hole symmetry in the auxiliary system is obtained when
\begin{align}
\label{egph}
 E_{ij} = (-1)^{i+j+1} E_{N_B+2-j,N_B+2-i} \,, \nonumber \\
 \Ga1_{ij} = (-1)^{i+j} \Ga2_{N_B+2-j,N_B+2-i}\,,
\end{align}
for 
$i,j \in\{1,\dots,N_B+1 \}$. 
More details for the particular form of $\vv{E}$, $\vgam1$ and $\vgam2$ are given below in App.~\ref{mat}.

\subsection{Interacting case}
\label{inte}
Despite 
\ea{
 the 
}
 fact that the solution of the interacting impurity problem is not the main topic of the present work, it is the main purpose of the overall approach. We thus briefly discuss here some relevant issues, in connection to the evaluation of particular observables  of the physical system from results of the auxiliary system. More details can be found in Refs.~\cite{do.nu.14,do.ga.15}

As already discussed, by  mapping onto an auxiliary {\em interacting} open quantum system of finite size described by the Lindblad equation~\eeqref{eq:Leq}, we obtain a many-body problem which can be solved exactly or at least with high numerical precision, provided $N_B$ is not too large. In \tcite{do.nu.14} we presented a solution strategy based on exact diagonalization (ED) with Krylov space methods, and in \tcite{do.ga.15} one based on matrix product states (MPS).
In the end both techniques allow us to determine the interacting impurity Green's function $\und G_{\aux,\iii}(\omega)$ of the auxiliary system. As discussed above, in the limit $\Daux\to\Dph$ (i.e. for large $N_B$) this becomes equivalent to the physical one $\und G_{\ph,\iii}(\omega)$. 
However, this equivalence only holds for impurity correlation functions, and, for example, it does not apply for the current 
flowing from a left 
($\lam=\lef$) to a   right ($\lam=\rig$) reservoir
across the impurity. 
Therefore,  the current  evaluated within the auxiliary Lindblad system does not 
\ea{
necessarily  
}
correspond to the physical current {\em even for large $N_B$}, unless one fits the  bath hybridisation functions
 $\und\Delta_{\ph,\lam}(\omega)$ for the left and right reservoirs
separately. Such a separate fit, however, is not necessary and would simply 
\ea{
worsen 
}
the overall accuracy 
for a given $N_B$. 
Once the approximate $\und G_{\ph,\iii}(\omega)\approx\und G_{\aux,\iii}(\omega)$ is known, 
the current  of the {\em physical system} 
  can be evaluated by means of the well-known
Meir-Wingreen expression~\cite{me.wi.92,ha.ja,jauh}, 
however, by using the Fermi functions and density of states (hybridisation functions) 
of the two {\em physical} reservoirs separately.
Therefore, the knowledge of $\und G_{\aux,\iii}(\omega)$ enables one to compute 
\ea{
most 
}
quantities of interest.

An additional step 
consists in extracting just the self-energy from the solution of the auxiliary impurity system
\[
 \und \Sigma_\aux(\omega) = \und G_\aux^{-1}(\omega) -  \und G_{\aux,\iii}^{-1}(\omega) \,.
\]
and inserting it into the Dyson equation for the physical system with the exact physical noninteracting Green's function
\beq
\label{gpi}
\und G_{\ph,\iii}(\omega) \approx \left(\und G_{\ph}(\omega)^{-1} -  \und \Sigma_\aux(\omega) \right)^{-1}\;.
\eeq
Clearly, this step
 is
only useful when the relation \eeqref{dphdaux} is approximate, since
for $\Daux\to\Dph$ also the noninteracting Green's functions
$\und G_{\ph}(\omega)$ and $\und G_{\aux}(\omega)$ would coincide, i.e. in the hypothetical $N_B \to \infty$ case, and one could just set $\und G_{\ph,\iii}(\omega) \to \und G_{\aux,\iii}(\omega)$.
For finite $N_B$
this substitution has the advantage that 
in \eqref{gpi} the noninteracting part $\und G_{\ph}(\omega)$ is exact, and 
the 
approximation
\eeqref{dphdaux} only affects
  the self energy.

\subsection{Different geometries for the auxiliary system}
\label{sec:setups}
\begin{figure*}
\begin{center}
\includegraphics[width=0.9\textwidth]{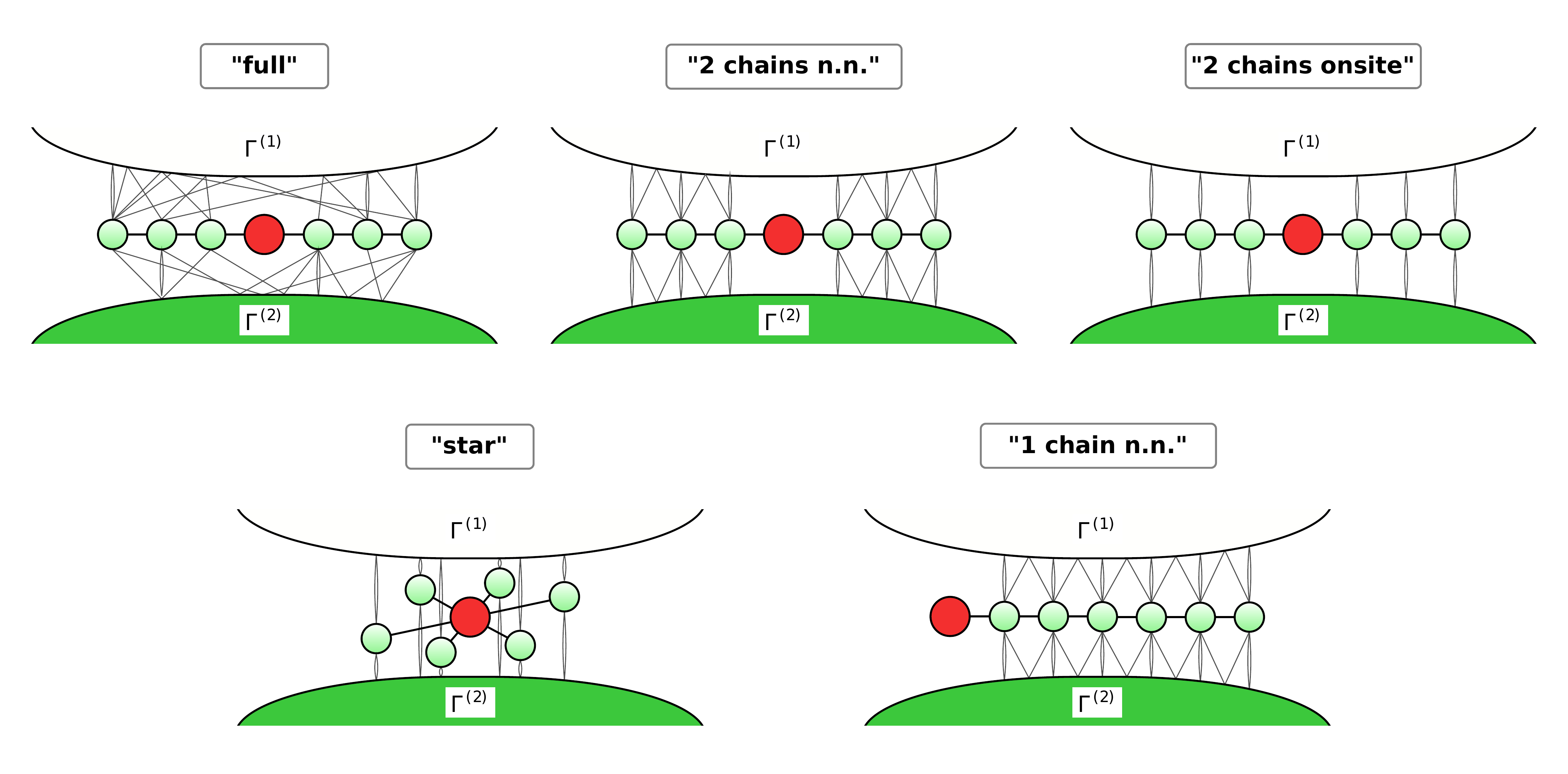}
\caption{
Sketch of the five geometries (setups) for the auxiliary system \eeqref{eq:Leq}. 
An explicit form of the corresponding matrices for $N_B=4$ is given in App.~\ref{mat}.
The impurity is represented by a red circle while the bath sites are filled green ones. The hoppings described by the matrix
 $\vv E$ are represented by thick black lines. The couplings to the Markovian environments given by $\vgam{1/2}$ are expressed by grey lines connected to empty ($\vv \Gamma^{(1)}$) or full ($\vv\Gamma^{(2)}$) reservoirs. 
On-site terms in the $\vgam{1/2}$-matrices are illustrated as a double grey line. 
The setup ``full'' represents the most general case with dense $\vgam1$ and $\vgam2$ matrices, which couple each bath site with every other one via the $\Gamma^{(1/2)}_{i,j}$.  
For simplicity, we don't depict  all terms for this ``full'' case.
 For the other (sparse) cases all couplings are drawn, and n.n. denotes nearest neighbor terms in $\vgam{1/2}$.
}
\label{fig:setups_sketch}
\end{center}
\vspace{-1em}
\end{figure*}

With the goal in mind of providing the best approximation to the full interacting impurity problem described by the Hamiltonian \eeqref{hfull}, we would like to  approximate $\Dph$ by $\Daux$ as accurately as possible for a given number of bath sites $N_B$.  
In principle, one has the freedom to choose different geometries for the auxiliary system and a generic set of five different setups 
is depicted in Fig.~\ref{fig:setups_sketch}. 
(An explicit form of the corresponding matrices for $N_B=4$ is given in App.~\ref{mat}.)
For large $N_B$ they all converge to the exact solution $\Daux\to\Dph$, the question is how fast. In \se\ref{sec:results} we want to elaborate on this point in detail and present results obtained with those geometries, which we briefly discuss and motivate here. 

In all cases one can restrict the geometries to a sparse
\ea{
 (e.g. tridiagonal) and real-valued matrix $\vv{E}$.   
}
As commonly true for impurity problems, the physics on the impurity site is invariant under unitary transformations among bath sites only. For an arbitrary unitary tranformation $\vv U$ 
with $U_{if}=U_{fi} =\delta_{if}$ 
 to new fermionic operators, one obtains an analogous auxiliary system with modified bath parameters $\vv{E}' = \vv U^\dagger \vv{E} \vv U$, $\vv{\Ga1}' = \vv U^\dagger \vv{\Ga1} \vv U$ and $\vv{\Ga2}' = \vv U^\dagger \vv{\Ga2} \vv U$. It is easy to check that the $ff$-component of the Green's functions \eqs{gro} and (\ref{gkeff}) is not affected by this transformation. 
Therefore, we choose without loss of generality $\vv{E}$ to be sparse as well as real, and for $\vgam{1/2}$ in the most general case dense matrices with $\mathcal{O}(N_B^2)$ 
 parameters. The particular form of $\vv{E}$ is irrelevant, i.e. whether it is diagonal for bath sites (star) or tridiagonal (chain), as long as the $\vgam{1/2}$ matrices are transformed accordingly. 

Such a general geometry with sparse $\vv{E}$ and dense $\vgam{1/2}$ is referred to as ``full'' setup in the following.
\ea{
Here, we will further distinguish between the case in which the $\vgam{1/2}$ are real or they have complex elements (``full complex''). 
}
In addition we consider the four sparse cases ``2 chains n.n.'', ``2 chains onsite'', ``star'', and ``1 chain n.n.'', in which 
also the $\vgam{1/2}$
 are sparse.
 The meaning of these abbreviations is given in \fig{fig:setups_sketch}, see also App.~\ref{mat}.  
These sparse geometries are however not linked to each other by unitary transformations and represent inequivalent subsets of the ``full'' setup. 
Which one of these is advantageous in practice is not obvious {\em a priori}, and discussed in the next 
\ea{
section.~\footnote{The number $C(N_B)$ of fit parameters for each geometry for the  
 particle-hole symmetric case, which we consider here,
is presented in App.~\ref{mat}.
}
}

\ea{
The ``full'' geometry 
comprises all other ones and thus, obviously,
gives the best possible fit for a given $N_B$. In addition, one can allow for the off diagonal matrix elements of the $\vgam{1/2}$ to be complex, thus extending the set of fit parameters. 
Nevertheless,
}
the sparse setups may be of great value for sophisticated manybody solution strategies for the interacting Lindblad equation, such as MPS. We made use of the ``full'' setup 
\ea{
(with real parameters)  
}
in the ED treatment \tcite{do.nu.14}, which is applicable to dense $\vv{\Ga1}$ and $\vv{\Ga2}$ matrices, and could consider up to $N_B=6$. 
Larger systems are prohibitive due to the exponentially increasing Hilbert space.~\footnoterecall{kdens}
 In the recent MPS implementation \tcite{do.ga.15}, on the contrary, we could consider as many as $N_B=16$ bath sites. However, in favour of the applicability of MPS methods one should avoid long-ranged hoppings and we thus employed the ``2 chains n.n.'' geometry. As becomes evident also from the results below, the gain in $N_B$ hereby outweighs the restriction of the fit setup, so that the MPS approach is clearly superior. Also the other sparse setups investigated below are possible candidates for MPS, see also \tcite{wo.mc.14}. Besides this, 
approaches such as 
the above mentioned buffer zone scheme and variations of it,~\cite{dz.ko.11sa,dz.ko.11,aj.ba.12} which are often applied concepts in Lindblad-type representation of noninteracting environments, are related to the ``star'' geometry, see also the discussion below.

\section{Results}
\label{sec:results}

\begin{figure*}
\begin{center}
\includegraphics[width=1\textwidth]{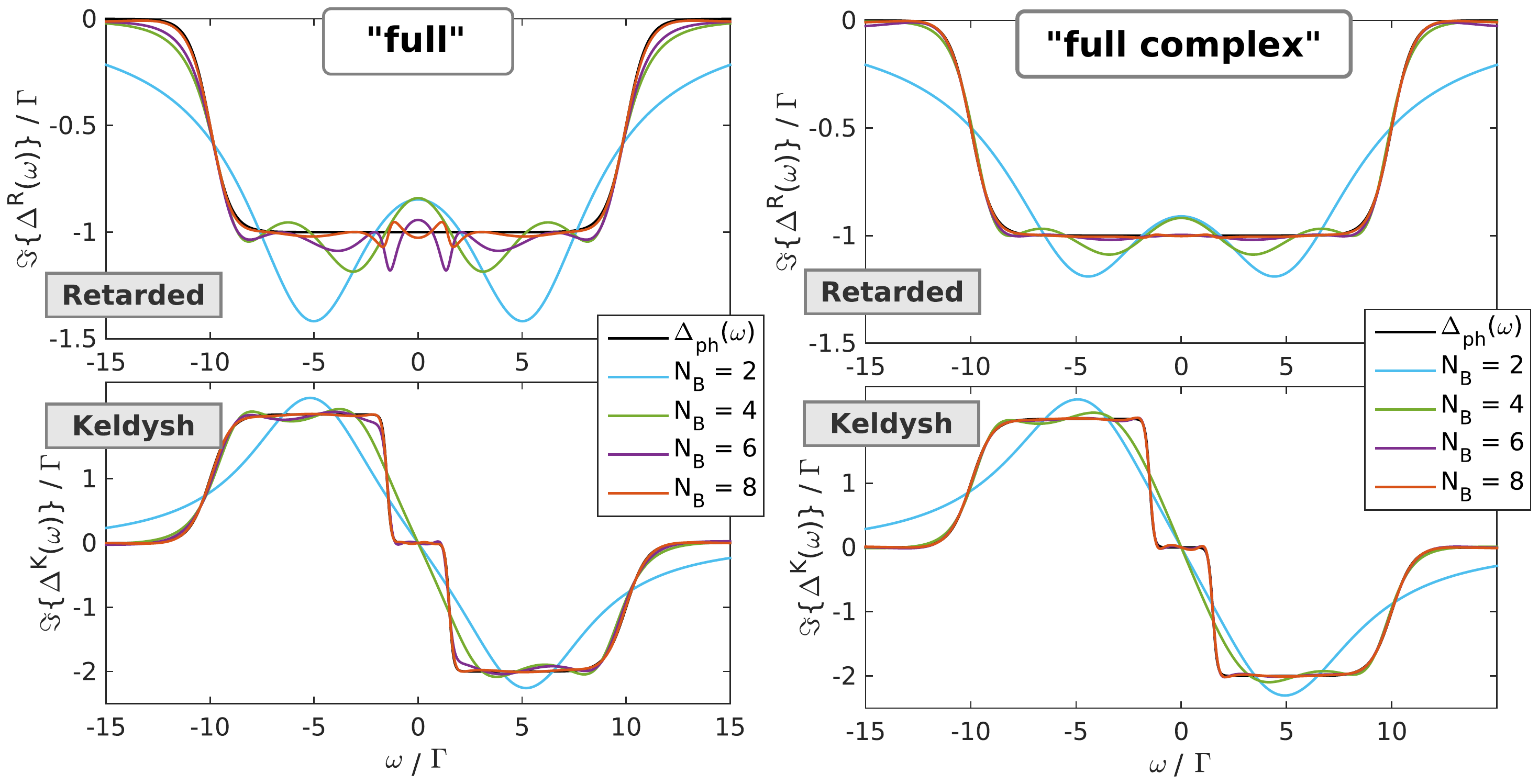}
\caption{
Fit to the bath hybridization functions for 
the ``full'' setups (real and complex) 
(see \fig{fig:setups_sketch}). The physical $\Dph$ (black lines) describes a reservoir with a flat 
density of states with hybridization strength $\Gamma$ and a half bandwidth of $D=10\,\Gamma$ which is smeared at the edges.
An applied bias voltage  $\phi = 3\,\Gamma$ shifts the chemical potentials of the two  reservoirs (leads) anti-symmetrically and a  temperature of $T = 0.1\,\Gamma$ is considered here.
}
\label{fits1}
\end{center}
\end{figure*}

\begin{figure*}
\begin{center}
\includegraphics[width=1\textwidth]{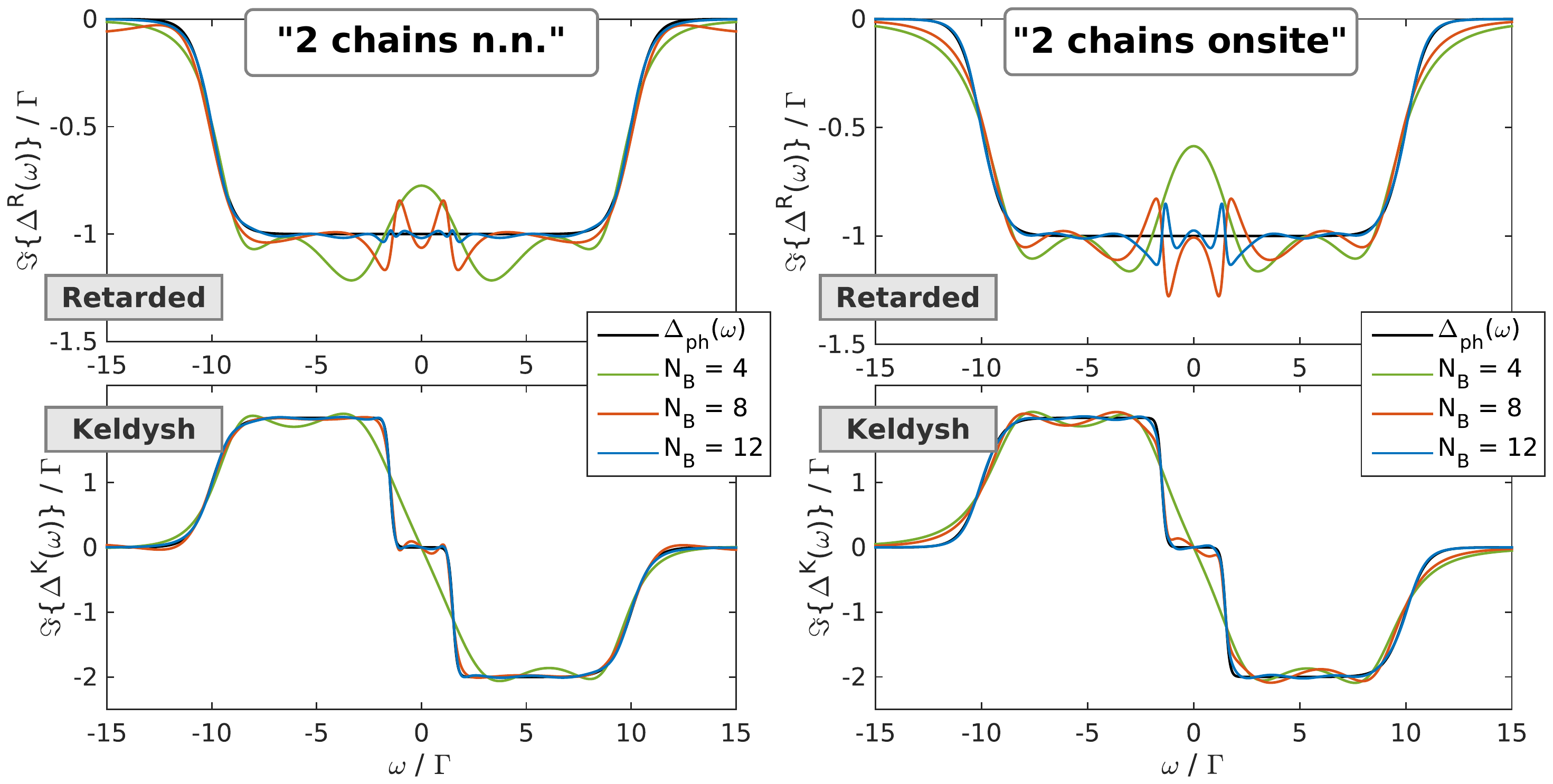}
\caption{
Same as Fig.~\ref{fits1} for the ``two-chains n.n.'' and  ``two-chains onsite'' 
setups.
}
\label{fits2}
\end{center}
\end{figure*}

\begin{figure*}
\begin{center}
\includegraphics[width=1\textwidth]{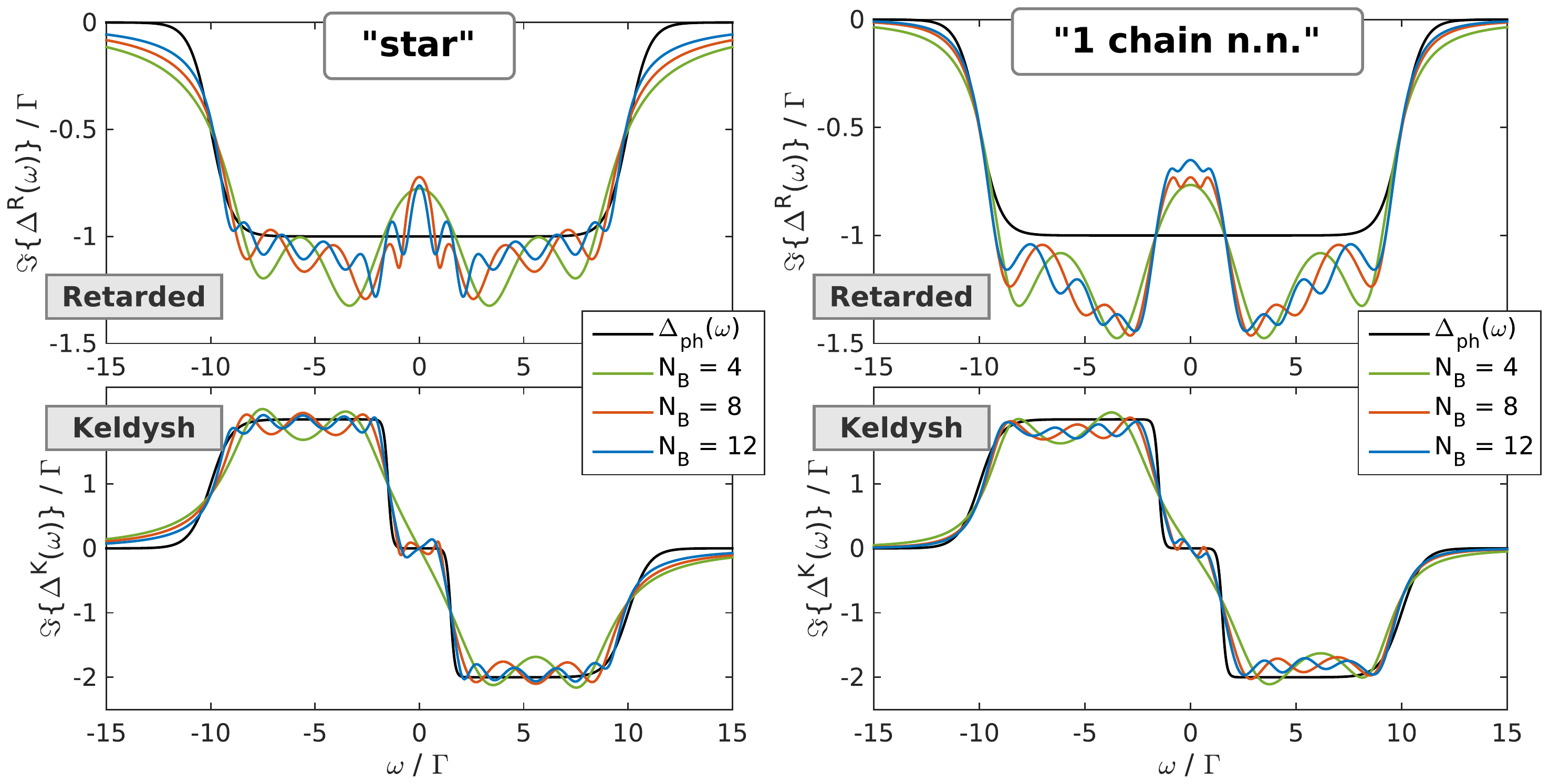}
\caption{
Same as Fig.~\ref{fits1} for the ``star'' and  ``1-chain n.n.'' 
setups.
}
\label{fits3}
\end{center}
\end{figure*}

As discussed above, while the ``full'' geometry is the most efficient one, for the purpose of employing efficient many-body 
eigenvalue solvers such as MPS, it is of great relevance to consider setups which feature only sparse $\vv E$, $\vgam1$ and $\vgam2$ matrices. 
Furthermore, it is also of general interest to investigate the importance of long-range terms in the $\vgam{1/2}$-matrices, 
and why they are crucial in order to improve the fit. 
These are the questions that are addressed in this section. Moreover,
we will analyze the rate of convergence as a function of $N_B$ 
for the different setups shown in \fig{fig:setups_sketch}, and for different temperatures of the physical system. 
The detailed knowledge of the convergence properties is important in order to be able to estimate whether certain systems can be accurately treated or not.

\begin{figure*}
\begin{center}
\includegraphics[width=0.95\textwidth]{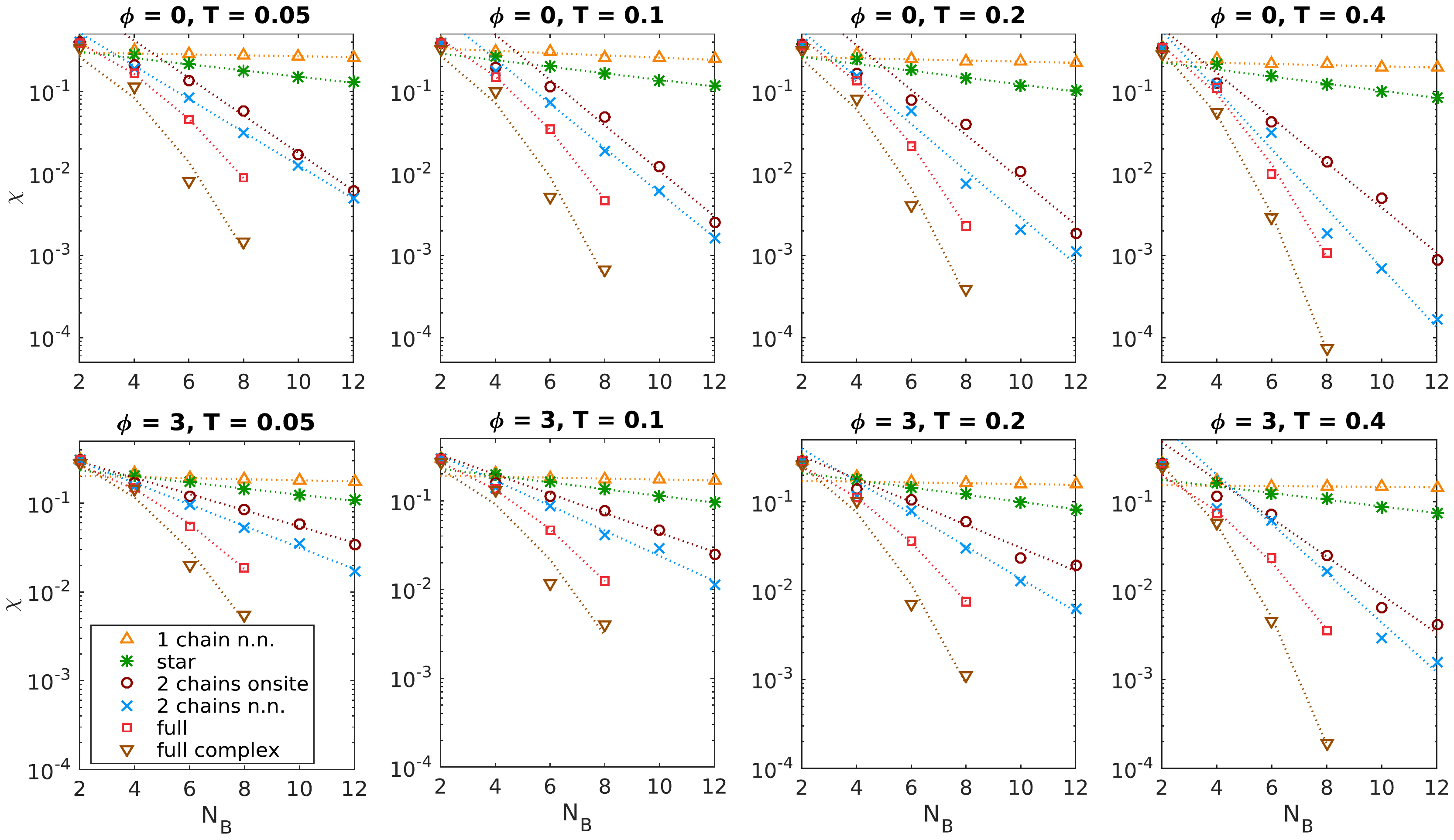}
\caption{
Minimal values of the cost function $\chi$, \eeqref{eq:costfunc}, as a function of the number $N_B$ of bath sites  for the 
\ea{
setups sketched in \fig{fig:setups_sketch} (including ``full complex'') 
}
for four temperatures 
$T = \{0.05\,\Gamma,\, 0.1\,\Gamma,\, 0.2\,\Gamma,\, 0.4\,\Gamma \}$ and two bias voltages $\phi=0$ and $\phi=3\,\Gamma$. 
Markers represent the raw data and dotted lines are  obtained from the fits of \fig{fig:conv_chi_dim}.}
\label{fig:conv_chi_NB}
\end{center}
\vspace{-1em}
\end{figure*}

We consider a physical system consisting of an impurity site coupled to two reservoirs (leads) at different chemical potentials, 
corresponding to a bias voltage $\phi$ across the impurity, and with a flat density of states as plotted in \figfits. 
Typical results for a given $\phi$ and temperature $T$ are shown in \figfits. 
\ea{
For the 
different setups
the quality of the fit  is measured by the minimum of the cost 
}
function \eeqref{eq:costfunc}. As discussed above, 
the ``full'' setups give the best results.
Already for a rather small number of bath sites $N_B\gtrsim 4$, a good agreement 
between $\und \Delta_\aux$ and $\und \Delta_\ph$ is achieved, and
the convergence is fast as a function of $N_B$. 
\ea{
Allowing for complex matrix elements produces a
 significant improvement.
The accuracy obtained with  $N_B=8$ for the real case 
is essentially achieved already with $N_B=6$ in the complex case (see also \fig{fig:conv_chi_NB}). 
Here, an excellent agreement is evident with minor differences in the Keldysh component. In the retarded component the largest differences occur at the positions of the jumps in the Keldysh component, i.e. at the chemical potentials.
This is a result of the simultaneous fit of the retarded and Keldysh components in \eeqref{eq:costfunc}, which produces  
  oscillations in the retarded component. 
These oscillations are strongly reduced in the complex case. 
By increasing the number of bath sites the amplitude  and the extension of these oscillations in the retarded component decay rapidly.

We now consider the sparse geometries.
In contrast to the ``full'' setups, no improvement is obtained by allowing the matrix elements to be complex in this case.
}
Among the sparse geometries, the ones with two chains are the most accurate. Both setups perform quite well. Again, a good agreement for small $N_B$ is obtained and a quick improvement shows up when increasing $N_B$. ``2 chains n.n.'' has off-diagonal $\vgam{1/2}$-terms in contrast to ``2 chains onsite'', which  leads to a faster convergence as seen e.g. for $N_B=12$. 
The ``star'' and most notably the ``1 chain n.n.'' geometry are clearly worse. Both exhibit a rather poor  convergence as a function of $N_B$.
 For the ``star'' setup, this is due to the fact that the fitted auxiliary hybridization function consists of  a sum of Lorentzian peaks. 
These enter in the Keldysh component with either positive or negative weights and can thus cancel each other. However, the rather broad Lorentzians with long $1/\omega^2$ tails make it apparently difficult to resolve the Fermi edges properly. The problem with slow convergence is most severe for the ``1 chain n.n.'' geometry. Here, the single chain is clearly inadequate to represent  at the same time the desired density of states and the sudden changes in the occupation number, see also the discussion below. While the Keldysh component is roughly reproduced,  this comes at the price of large oscillations in the retarded one. In addition,  the improvements with increasing $N_B$ are minor and the results for $N_B=4$ and $N_B=12$ are very close to each other.

\begin{figure*}
\begin{center}
\includegraphics[width=0.95\textwidth]{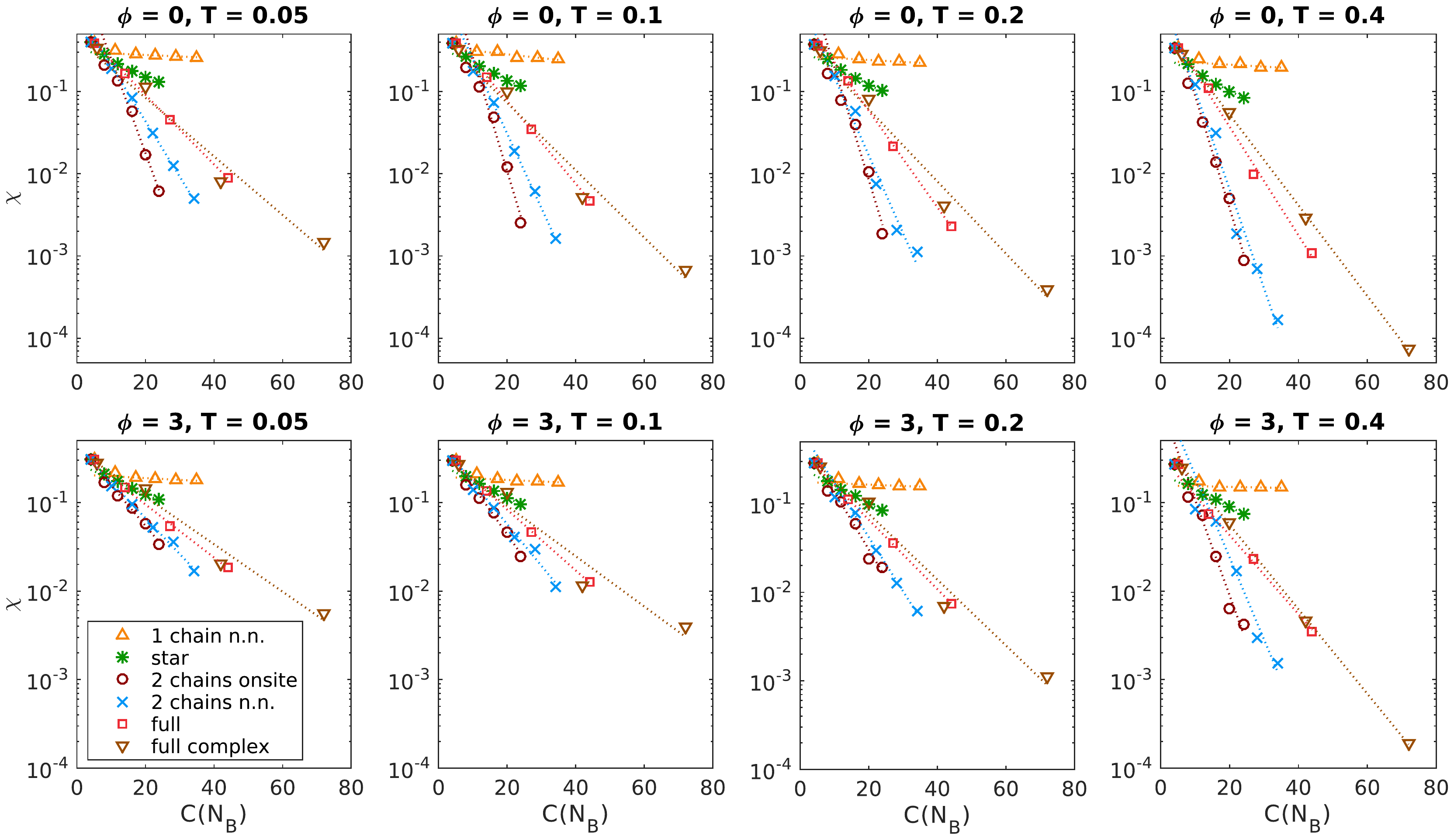}
\caption{
Same as figure 
\fig{fig:conv_chi_NB} but plotted versus the number of fit parameters $C(N_B)$.
In order to resolve the scaling with temperature more reliably, we exclude
the two data points with the smallest $N_B$ from  each of the linear fits, which have not enough structures to 
resolve low-energy scales.
Dotted lines represent results of linear fits in these semi-logarithmic plots. The 
temperature dependence of the convergence rates 
\ea{
(as a function of $N_B$)  
}
obtained in this way are illustrated in \fig{fig:conv_rates}.  
}
\label{fig:conv_chi_dim}
\end{center}
\vspace{-1em}
\end{figure*}
\begin{figure*}
\begin{center}
\includegraphics[width=0.7\textwidth]{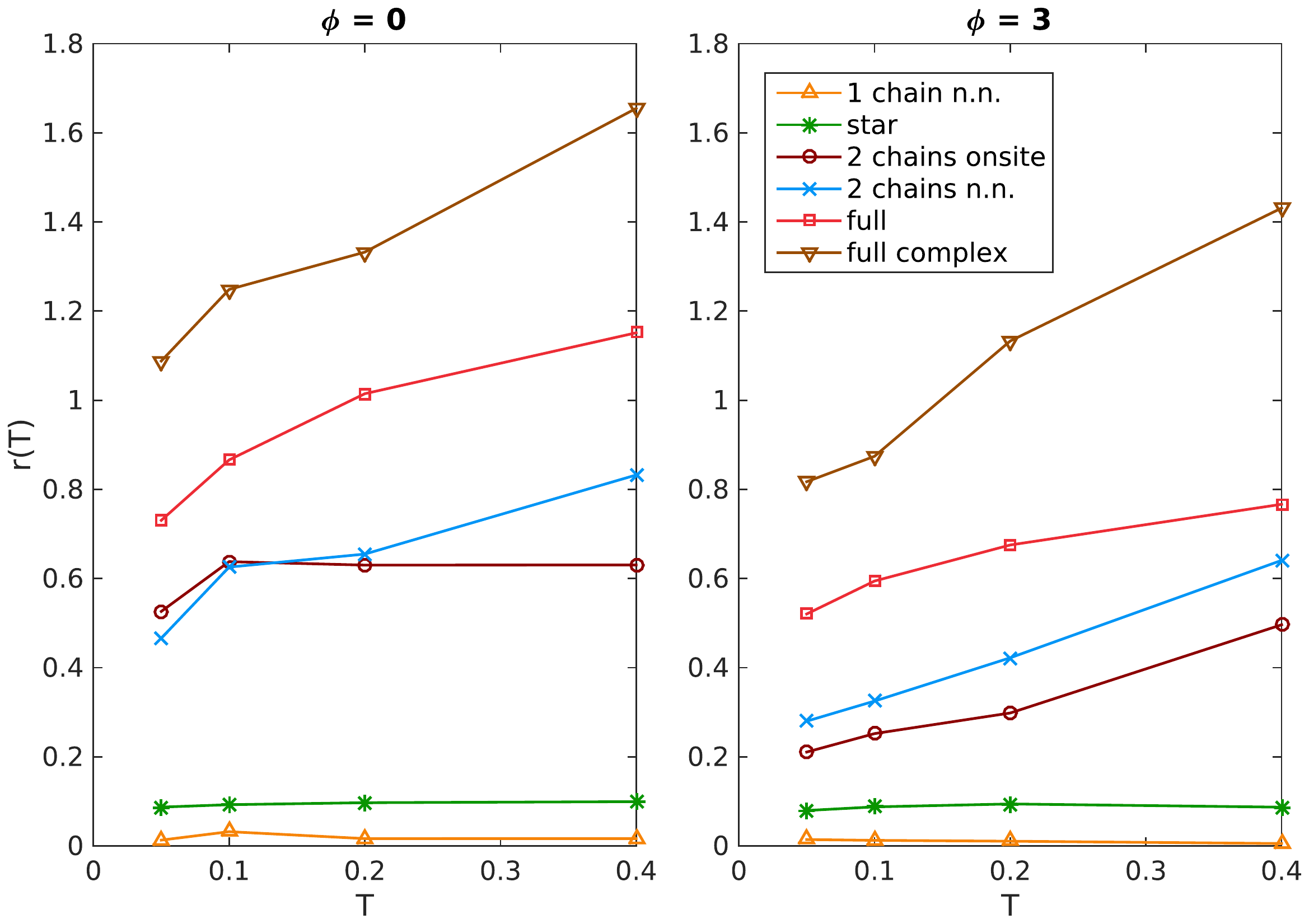}
\caption{Estimated convergence rates  obtained 
from the data in \fig{fig:conv_chi_NB} plotted
as a function of temperature. The rates for the sparse setups are obtained by assuming
 $\chi \propto \exp[-r(T)N_B]$. For the ``full'' setups, the exponent is quadratic in $N_B$, therefore we plotted the differential rate, defined as
$-\frac{d \log\chi}{d N_B}$ evaluated at $N_B=6$.
}
 \label{fig:conv_rates}
\end{center}
\vspace{-1em}
\end{figure*}

The behavior just discussed is even more visible in the convergence study presented in \fig{fig:conv_chi_NB} and \fig{fig:conv_chi_dim}. 
In \fig{fig:conv_chi_NB} the minimal values of the cost function $\chi$, \eeqref{eq:costfunc}, for various values of $N_B$ and the different setups are shown. Four different temperatures and each of them with $\phi=0$ and $\phi=3\,\Gamma$ are considered.
\ea{
As expected, the ``full complex'' setup gives the lowest values of $\chi$ in all cases, and, moreover, the fastest rate of convergence as a function of $N_B$. The ``full''  setup, without complex terms also performs quite well.
}
The sparse geometries ``2 chains n.n.'' and ``2 chains onsite'' perform not as well, which is not surprising since only restricted subsets
of the full available fit parameters are used in this case. Nevertheless, these setups achieve a rather high rate of convergence. This shows that
of all possible geometries, ``2 chains'' ones apparently contain the most relevant contributions. In most cases studied here, the off-diagonal $\vgam{1/2}$-terms in ``2 chains n.n.'' result in a significant improvement compared to ``2 chains onsite'', which is the reason why we  favored the former in our MPS many-body calculations performed in \tcite{do.ga.15}. In that work we found that an accuracy of at least $\chi\approx 10^{-2}$ was necessary in order to properly account for Kondo physics. This could  be reached already for $N_B \approx 12$.

We now discuss the ``star'' setup. In order to present a fair comparison with the other geometries
we optimize all available parameters within this geometry, namely all $E_{i,i}$, $E_{i,f}$, and $\Gamma^{(1/2)}_{i,i}$. 
In this way we obtain an exponential convergence as for the other setups, although with a significantly smaller rate.
\ea{
One  
}
should note that in standard  buffer zone approaches~\cite{dz.ko.11sa,dz.ko.11,aj.ba.12}
an equidistant energy spacing $\Delta \epsilon_i \approx 2D/N_B$ with equal onsite $\vgam{1/2}$-terms is often assumed for the bath sites.
 Clearly, such a discretization approach cannot converge exponentially and it is only first-order accurate in the spacing $\Delta \epsilon_i$. 
Therefore, the value of the cost function presented here for the ``star'' setup can be seen as a lower bound for the buffer zone approach. Despite of the exponential convergence of the ``star'' geometry, it becomes apparent from \fig{fig:conv_chi_NB} that a very slow rate of convergence is achieved. To reach an accuracy $\chi\approx 10^{-2}$ for the case $T = 0.05\,\Gamma$ and $\phi=0$ for instance, much larger auxiliary systems with $N_B \approx 40$ would be needed. 
For the  MPS-solver used in \tcite{do.ga.15} such large auxiliary systems are clearly out of reach.
Therefore, the present analysis clearly demonstrates the huge advantage of optimizing the bath parameters of the auxiliary system,
and furthermore, of choosing an appropriate geometry when considering only a restricted subset of the ``full'' setup.

Let us now turn to the results for the ``1 chain n.n.'' setup in \fig{fig:conv_chi_NB}. 
Despite of the poor performance and the strongly limited practical use, the observed behavior is interesting from a fundamental point of view. As becomes evident from the results, a single chain with local dissipators is a particularly bad choice in order to represent a partially filled bath. The  convergence is very slow and an extremely long chain would be needed in order to achieve results comparable to the other geometries.
As shown above, a drastic improvement is obtained when using two chains instead. This would be more or less intuitive for 
the nonequilibrium case, in which the physical system also consists of two baths. However, the advantage of the ``2 chains'' geometry over the ``1 chain'' case is even more pronounced in the equilibrium case (see $\Phi=0$). Another important observation to better understand this is the following: In \tcite{do.ga.15} we found nearly identical accuracies when considering the ``2 chains'' geometry as used here, or a filled/empty restriction of it. In the latter case one chain has the purpose of representing the filled spectrum and the other chain the empty spectrum of the physical hybridization function~\footnote{The filled (empty) spectrum corresponds to the lesser (greater) hybridization function $\Delta^{<}$ ($\Delta^{>}$), and furthermore: $\Delta^{</>} = \Delta^{K}/2 \mp i\, \iim\{ \Delta^{R}  \}$.}, and not necessarily the two physical reservoirs. 
This shows that a single chain of small size is very well-suited to reproduce a certain density of states, but not simultaneously a Fermi edge or other sharp 
\ea{
changes in the occupation number. Furthermore, 
}
 a ``2 chains'' filled/empty setup seems to be a rather natural representation which contains the most relevant subset of the ``full'' geometry. Here, the resolution of sharp features in $\Dph$, which either correspond to band edges or to sudden occupation changes at the Fermi edges, are resolved by appropriate Hermitian couplings $\vv E$ and correspondings broadenings/couplings stemming from $\vgam{1/2}$. In this way, the filled and empty chain together can well reproduce sharp features in $\DRph$ and $\DKph$.\footnote{
From this point of view, the additional improvement in the ``full'' setups can roughly be interpreted in such a way that one achieves an optimal linear combination of filled/empty states with the long-ranged couplings in $\vgam{1/2}$.   
}

Additionally to the convergence as a function of $N_B$ we depict in \fig{fig:conv_chi_dim} the cost function versus the number of available fit parameters $C(N_B)$. As can be seen, the trends in the semi-logarithmic plot are well described by straight lines in all cases, which clearly shows the achieved exponential convergence with respect to $C(N_B)$. For the sparse setups this means that $\chi \propto \exp[-\mathcal{O}(N_B)]$ whereas for the ``full'' 
\ea{
setups 
even $\chi \propto \exp[-\mathcal{O}(N_B^2)]$. Due to this, the ``full''
 geometries converge
}
 much quicker, as observed in the results above. With respect to the number of fit parameters, however, the ``2 chains'' setups perform best. Again, this signifies that these setups contain the most relevant subset of all possible fit parameters.

Another important aspect is the dependence of the convergence rate $r(T)$ on temperature.
The estimated rates $r(T)$ for each setup are depicted in \fig{fig:conv_rates}. Of course, the superior scaling of the ``full'' and the ``2 chains'' setups is also apparent in the magnitude of $r(T)$. Furthermore, in all cases one observes the trend that the higher the temperature the faster the convergence. This can be understood from the fact that at high $T$ the Keldysh component $\DKph$ is weakly $\omega$-dependent so that less bath sites are necessary for a reliable fit. Eventually, in the $T\to\infty$ and wide-band limit the Markov approximation becomes even exact.
In the other extreme $T\to0$ limit, discontinuous functions are present in $\DKph$, produced by the abrupt Fermi edges. However, each of the frequency dependent functions in the effective set given by Eqs.~(\ref{gro}-\ref{eq:DKaux}) is continuous. Therefore, $T\to0$ can only be reproduced in the limit $N_B\to\infty$. This explains the observed trend that, for a given $N_B$,  
the high-temperature regime is generally better represented than the low-temperature one. 
Furthermore, a nonzero $\phi$ tends to result in larger values for the cost function, see also \fig{fig:conv_chi_NB}. 
\footnote{Note that the difficulty of the fit, i.e. the magnitude of $\chi$, is determined by the degree of variations in $\Delta_\ph$ and the length-scale of these variations wrt. the half bandwidth $D$. The coupling strength $\Gamma$ of the leads enters only trivially. Therefore $\phi/D$ and $T/D$ determine $\chi$.
}

\subsection{Discussion of further aspects}
The 
\ea{
present 
}
approach is equally suitable  to describe a system in equilibrium as well as out of equilibrium. In the first case it 
becomes competitive 
with  conventional 
ED- and MPS-impurity solvers for DMFT based on a bath without Lindblad terms.
The distinction between the equilibrium or nonequilibrium situation shows up in 
 the properties of the  bath hybridisation function $\Dph$. 
 In the equilibrium case its  Keldysh and retarded part will fullfill the fluctuation-dissipation theorem. 
Interestingly, the equilibrium problem
is mapped onto an auxiliary nonequilibrium one, since
 a current will typically flow from $\Ga2$ to $\Ga1$ dissipators. An example is the case discussed in Sec.~\ref{sec:results}
of a two-chain geometry with a completely  empty and a full one.
 Such  geometry can be used to describe an equilibrium situation {\em at the impurity} as well, as long as 
$\DKaux$ and $\DRaux$ are chosen to fullfill the fluctuation-dissipation theorem. Nevertheless, a current will flow 
from one chain to the other across the impurity, which, however, will be in equilibrium. 
Notice that since the mapping will be approximate for finite $N_B$, there will be small deviations from 
the fluctuation-dissipation theorem.

Another interesting aspect is the role of chemical potential(s) $\mu_\al$ and temperature(s) $T_\al$ of the different physical reservoirs.
These determine only indirectly the values of the parameters of the auxiliary system
$\vv E$, $\vgam1$, $\vgam2$. 
More specifically
$\mu_\al$ and  $T_\al$ first determine $\DKph$ via \eeqref{delij} and \eeqref{gkal}. In a second step, via the requirement  \eeqref{dphdaux}
and the corresponding fit procedure, they finally determine the auxiliary parameters. 
For low temperatures, the chemical potentials $\mu_\al$ will then appear as sharp changes in $\DKph$.
This is in contrast to more direct  approaches, such as buffer-zone based ones, in which the parameters are directly determined form
$\mu_\al$ and  $T_\al$, in equations such as  \eeqref{li1g} for each bath level. 
Both methods have their advantages: 
Direct approaches can be more convenient, for example in NRG~\cite{sc.go.16u}. On the other hand, a fit procedure like the present one produces a much faster, exponential,  convergence.

\section{Summary and conclusions}
\label{summ}
In this work, a scheme for mapping the hybridization function of correlated quantum impurity problems with non-Markovian fermionic reservoirs onto an auxiliary open quantum system was 
\ea{
developed and 
}
presented in a general framework and discussed in detail. The approach as outlined here can be used to model transport through interacting impurities, Hubbard chains or small clusters and molecules. The key aspect is to replace the infinite fermionic reservoirs of the original problem by a combination of a small number $N_B$ of bath levels plus Markovian terms. By this we arrive at a finite open quantum system described by a Lindblad equation, whose manybody problem can be solved with high accuracy by numerical techniques. However, despite of the Markovian environments for the bath levels the thereby approximated hybridization function is clearly {\em non-Markovian} 
\ea{
at the impurity site in the sense that it has a frequency dependence, which is a consequence of the  memory effects of the environment.
}
While this idea is not new, the key point of our work is the formulation of
an optimization procedure in order  to determine the parameters of the auxiliary bath levels. This allows us to achieve an exponential convergence of the mapping, as clearly demonstrated in this work.

In the mapping one has certain degrees of freedom and different geometries for the auxiliary system are possible. In this work we discussed a variety of choices in detail and compared their performance. 
\ea{
When using Krylov-based many-body approaches it is convenient to take advantage of  as many fit parameters as possible. For these cases the ``full'' setups are the best choice.
Here we also showed that further allowing for complex matrix elements (``full complex'') 
drastically improved 
the accuracy of the mapping
with respect to the plain real ``full'' setup, which we used in Ref.~\onlinecite{do.nu.14}.
}
With efficient manybody solution techniques, such as MPS, in mind, it is of advantage to restrict the auxiliary quantum system to a sparse form. For this we analyzed four sparse
\ea{
 setups. 
}
 The results revealed the most relevant degrees of freedom in the auxiliary system and demonstrated clearly that the performance of different sparse setups may differ by orders of magnitude. In particular, the well-known ``star'' geometry turned out to exhibit a very slow rate of convergence when increasing $N_B$, and also a geometry with ``1 chain'' and local Lindblad drivings performed much worse than the other cases. In contrast, setups with ``2 chains'' and local Lindblad drivings yielded very good results, with an accuracy orders of magnitude below the other two sparse cases. With this knowledge it is possible, on the one hand, to employ efficient sparse setups which yield already for modest values of $N_B$ very accurate results, and on the other hand, one can better understand the underlying mechanisms of the mapping. Together with the findings mentioned in \tcite{do.ga.15}, we can conclude that a so-called filled/empty geometry with ``2 chains'' is essentially a natural representation of a non-Markovian reservoir by auxiliary Lindblad levels. In this geometry one chain has the purpose of reproducing the filled spectrum of the original reservoir whereas the other chain the empty spectrum. This is achieved in each chain separately by an optimal combination of hoppings between the bath levels and couplings to {\em one} Markovian environment, which is then either completely filled or empty. By this separation it is possible to resolve sharp features in the original hybridization function in great detail, which may correspond to  sudden occupation changes at the Fermi edges or  band borders. A single chain coupled to filled and empty Markovian environments, on the contrary, cannot simultaneously represent a
\ea{
particular 
}
density of states and a partially filled spectrum appropriately, as evident from the ``1 chain'' setup.

Besides comparing different auxiliary setups to each other, we also analyzed the general convergence properties in detail. As mentioned above, a clear exponential convergence was found in all cases, which can be accounted to the optimization strategy for the bath parameters. Furthermore, a generic set of equilibrium and nonequilibrium reservoirs with various temperatures was chosen for the original system. From this we found the expected trend for all setups that the high-temperature regime is better represented by the auxiliary system than the low-temperature one, i.e. the rate of convergence of the mapping increases with temperature. Therefore, to achieve a given accuracy it is more challenging to resolve low temperatures and larger auxiliary systems must be considered. 
The plain exponential convergence shown here yields  a simple tool to extrapolate results for low $N_B$ to higher values, and by this to judge the feasability of treating certain physical situations.

While we did not discuss this in the present work, it would be probably useful to exploit the freedom in the choice  of the cost function~\eeqref{eq:costfunc}, 
and in particular, of the corresponding weight function $W(\omega)$. For example, one could increase the weight  around the chemical potential in order, possibly, to achieve a better resolution at  low energies. An even more appropriate strategy in this sense would be to
 combine the present approach to NRG ideas such as the logarithmic discretisation, work along these lines is in progress, see also Ref.~\onlinecite{sc.go.16u}.
 \eeqref{eq:costfunc} strongly disfavours delta-function-like or strongly oscillating spectra. In some cases, such oscillations may be unimportant, especially if they occur at high energies.  Therefore, it would be useful to adopt a cost function which does not penalize them. This could be achieved, for example, by
convoluting $\Dph$ and $\Daux$ with a suitable 
 ``smoothing'' function before evaluating their difference ~\eeqref{eq:costfunc}. Alternatively, one could use cost functions based on
differences in spectral moments up to a certain order.

In this work we considered the simplest case of a single impurity Anderson model (SIAM) in order to focus on the mapping itself. To treat more extended cluster or multi-level problems essentially the same approach can be used. In the simple case of diagonal cluster hybridization functions exactly the same equations are applicable to model each reservoir separately by auxiliary Lindblad levels. But also non-diagonal hybridization functions can be treated, of course. For this purpose the approach was presented here in a more general framework before focusing particularly on the SIAM.

Besides the technical aspects, we believe that the presented study contains relevant information to the general question of the representability of {\em non-Markovian} fermionic reservoirs by open quantum systems, and in particular by Lindblad-type equations. We expect that the insights gained in this work may contribute also to other closely related fields on Markovian and non-Markovian quantum master equations.

\section*{Acknowlegments}
We would like to thank H.G. Evertz, M. Nuss, F. Schwarz, J. von Delft and A. Weichselbaum for fruitful discussions.
This work was
partlially supported by the Austrian Science Fund (FWF) P26508, and P24081, and NaWi Graz.
The calculations were partly performed on the D-Cluster Graz 
and on the VSC-3 cluster Vienna

\appendix

\section{Multi-dimensional minimization}
\label{app:PT}
In this section we provide detailed informations for readers interested in an actual implementation. Furthermore, a working code is available on request. To obtain it 
\ea{just}
 contact us via e-mail. Much of the information below is contained in standard textbooks and reviews. 
However, for completeness we outline here the standard algorithm in detail
and point out  choices we made, which turned out to be convenient for the specific problem.

As stated above, a single evaluation of 
Eqs.~(\ref{gro}-\ref{eq:DKaux})
 is rather cheap numericaly
since it involves only one matrix inversion and 
multiplications of matrix of size $(N_B+1)\times(N_{B}+1)$
. Thus, the increase in computation time with $N_B$ is rather moderate. However, the multi-dimensional optimization problem itself is demanding and it strongly depends on the particular behavior of $\chi(\vv{x})$ when varying the set of parameters $\vv{x}$. In the worst case scenario, when $\chi(\vv{x})$ is a rough potential landscape with many local minima and short-scaled variations, one could imagine that it becomes necessary to nearly explore the whole parameter space. However, $\vv{x}$ is a continuous vector and even when assuming a fixed number of discrete values for each component in $\vv{x}$, one faces a number of points in parameter space that grows exponentially with $\dim(\vv{x})$. In the other extreme, for the case that $\chi(\vv{x})$ is quadratic in $\vv{x}$ it is well-known that a conjugate gradient scheme leads to the exact minimum in $\dim(\vv{x})$ iterations. What we found in practice, when performing the minimization within 
AMEA,~\cite{do.nu.14,do.ga.15}
 is that we have an intermediate situation which exhibits local minima, but gradient-based methods still work fine especially for smaller values of $N_B$. In the first work on the ED-solver, \tcite{do.nu.14}, we employed a quasi-Newton line search with many random starting points. This is particularly
 useful for $N_B<6$. However, the necessary number of starting points increases rapidly with $N_B$. Therefore, in the course of the work on the MPS-solver, \tcite{do.ga.15}, 
we looked for more efficient solution strategies. In the end we implemented a parallel tempering (PT) approach with feedback optimization, which is a Monte Carlo scheme that is able to overcome local minima. We describe it  in the following in more detail. In this way,  the minimization problem for the ED-solver with $N_B=6$ and for the MPS-solver with up to $N_B=16$ can be solved in reasonable time. This  amounts to minimizing in a space of  $\approx 30-60$ parameters in both geometries, depending on whether one has particle-hole symmetry  or not.%
\footnote{In order to perform the mapping for even larger systems efficiently it may be of interest to combine the PT approach with gradient based methods, for instance.}

\subsection{Markov chain Monte Carlo}
The PT algorithm is outlined in detail below. 
For completeness, let us first briefly recap the basic ideas of the underlying Markov chain Monte Carlo (MCMC), and  of the related simulated annealing algorithm.

MCMC techniques were originally developed to evaluate thermodynamic properties of classical systems which exhibit a very large phase space where simple sampling strategies fail. For our purposes here, we are interested in minimizing the cost function $\chi(\vv{x})$ as defined in \eeqref{eq:costfunc} with respect to the parameter vector $\vv x$. For such high-dimensional minimization problems one can adapt MCMC schemes by viewing $\chi(\vv{x})$ as an artificial energy and by introducing an artificial inverse temperature $\beta$. In the so-called simulated annealing one samples from the Boltzmann distribution $p(\vv{x}) = 1/Z\exp\left(-\chi(\vv{x})\beta\right)$ at a certain $\beta$, and then successively cools down the artificial temperature. Motivated by the behavior of true physical systems one expects to end up in the low-energy state when letting the system equilibrate and when cooling sufficiently slowly. Analogous to thermodynamics one can calculate the specific heat $C_H = \beta^2\braket{\Delta \chi(\vv{x})^2}$ and by this locate regions with large changes, i.e. phase transitions, where a slow cooling is critical. However, in practice it may be time consuming to realize the equilibration and sufficiently slow cooling, and for tests within AMEA we often ended up in local minima. In order to obtain a robust algorithm, which can also start from previous solutions as needed for instance within DMFT, we sought for a method which is able to efficiently overcome local minima and still systematically targets the low-energy states. For this a multicanonical and a PT algorithm were tested, whereby the latter turned out to be more convenient. In the following we briefly outline the PT scheme used within AMEA, and refer to \tcite{ever.09u,berg.05,berg.00,hu.ne.96,ea.de.05} for a thorough introduction to MCMC, simulated annealing, multicanonical sampling and PT.

As just stated, in a MCMC scheme one typically samples from the Boltzmann distribution $p(\vv{x}) = 1/Z\exp\left(-\chi(\vv{x})\beta\right)$ at some chosen inverse temperature $\beta$. This is done through an iteratively created chain of states $\{\vv x_l\}$, whereby  one avoids the explicit calculation of the partition function $Z$. An effective and well-known scheme for this is the Metropolis-Hastings algorithm~\cite{ever.09u,berg.05}. One starts out with some state $\vv x_l$ and proposes a new configuration $\vv x_k$, whereby it has to be ensured that every state of the system can be reached in order to achieve ergodicity.
The proposed state $\vv x_k$ is accepted with probability%
\footnote{In principle one has to take the proposal probabilities $q_{k,l}$ and $q_{l,k}$ into account. However, since we only consider the case $q_{k,l} = q_{l,k}$ here, the terms drop out of the equations and are neglected everywhere.}%
~\cite{ever.09u,berg.05}
\begin{equation}
 p_\mathrm{pacc.}^{l,k}  = \mathrm{min}\left\{ 1, \frac{p(\vv x_k)}{p(\vv x_l)} \right\} = \mathrm{min}\left\{ 1, e^{-(\chi(\vv x_k)-\chi(\vv x_l))\beta} \right\} \,.
 \label{eq:MCMCpacc}
\end{equation}
If the proposed configuration is accepted, then the next element $\vv x_{l+1}$ in the chain is $\vv x_k$, otherwise $\vv x_l$ again. From \eeqref{eq:MCMCpacc} it is obvious that $p_\mathrm{pacc.}^{l,k} = 1$ when $p(\vv x_k) > p(\vv x_l)$, so that an importance sampling towards regions where $p(\vv{x})$ is large is achieved. One can show that the algorithm fulfills detailed balance and draws a 
set of samples $\{\vv x_l\}$ that follow the desired distribution $p(\vv{x})$. However, stemming from the iterative construction, correlations in the chain are present which require a careful analysis for the purpose of statistical physics~\cite{ever.09u,berg.05}. For optimization problems, on the other hand, the situation is much simpler and one is just interested in the element in $\{\vv x_l\}$ which minimizes $\chi(\vv{x})$. Since a proposed step with $\chi(\vv x_k) < \chi(\vv x_l)$ is always accepted the algorithm targets minima, however, also uphill moves in configuration space are allowed with a probability depending exponentially on the barrier height $\Delta \chi_{k,l} = \chi(\vv x_k)-\chi(\vv x_l)$ and $\beta$. Effectively, uphill moves take only place when $\Delta \chi_{k,l} \beta \lesssim \mathcal{O}(1)$. For small values of $\beta$ large moves in configuration space with large $\Delta \chi_{k,l}$ are likely to be accepted, whereas for large $\beta$ the distribution $p(\vv{x})$ is peaked at minima in $\chi(\vv{x})$, so that especially those regions are sampled. For the latter case configurations in the chain $\{\vv x_l\}$ are generally more correlated and once a $\vv x_l$ corresponds to a local minimum the algorithm may stay there for very long times.

One has great freedom in defining a proposal distribution from which the new state $\vv x_k$ is drawn given the current configuration $\vv x_l$.%
\footnote{Note that for minimization purposes only one has in general the flexibility in designing the algorithm and the Boltzmann distribution or detailed balance are not compulsory.} %
Common choices are, for instance, a Gaussian or a Lorentzian distribution with the vector difference $\vv x_k - \vv x_l$ as argument. We favored the former  and updated each component $i$ with a probability according to~\cite{ever.09u}
\begin{equation}
 q_{l,k}^i = \frac{1}{\sqrt{2\pi}\sigma_i} e^{-\frac{(\vv x_k - \vv x_l)_i^2}{2\sigma_i^2}}\,. 
 \label{eq:MCMC_proposal}
\end{equation}
Hereby, a different step size $\sigma_i$ for each component is expedient since the potential landscape $\chi(\vv{x})$ around $\vv x_l$ is typically highly anisotropic. Ideally, one should make use of the covariance matrix $\Sigma_l$ of $\chi(\vv x_l)$ and consider as argument for the Gaussian instead $(\vv x_k - \vv x_l)^T \Sigma_l^{-1} (\vv x_k - \vv x_l)$~\cite{ever.09u}. However, we encountered the problem that the estimation of the covariance matrix at run time was strongly affected by noise and thus not feasible. The adjustment of the step sizes $\sigma_i$, on the contrary, can be done after a short number of updates by demanding that a value of $p_\mathrm{pacc.}^{l,k} \approx 0.5$ is reached on average when modifying the component $i$. For this we implemented a check at every single proposal, that increases $\sigma_i \rightarrow 1.1\, \sigma_i$ when $p_\mathrm{pacc.}^{l,k} > 0.6$ and decreases $\sigma_i \rightarrow 0.9\, \sigma_i$ when $p_\mathrm{pacc.}^{l,k} < 0.4$. Analogous to the treatment of spin systems, we define one sweep as a single update of all the components of $\vv x$.
\footnote{Again, different choices are possible. For instance in cases where $\dim (\vv x)$ is very large, random updates of the most relevant components could be more appropriate.}

\subsection{Parallel tempering}
In a PT algorithm one considers instead of sampling at one certain temperature a set of different temperatures $\beta_m^{-1}$ and corresponding replicas $\vv x_l^m$, each of which is evolved through a Markov chain. The largest $\beta_m$ thereby target local minima whereas low $\beta_m$ values allow for large moves in configuration space. The key idea of the PT approach is to let the individual replicas evolve dynamically in the set of $\beta_m$. By this one achieves that a replica at high $\beta_m$ values systematically targets local minima but can overcome potential barriers again when its inverse temperature is changed to lower values. As a result, the time scales to reach an absolute minimum are drastically reduced and an efficient sampling of the low-energy states is achieved. For the purpose of calculating thermodynamic properties one usually chooses a Metropolis-Hastings probability to swap two replicas with adjacent temperatures~\cite{hu.ne.96,ea.de.05}
\begin{align}
 p_{\mathrm{swap},l}^{m,m+1}  &= \mathrm{min}\left\{ 1, \frac{p^m(\vv x^{m+1}_l) p^{m+1}(\vv x^{m}_l)}{p^{m+1}(\vv x^{m+1}_l) p^{m}(\vv x^{m}_l)} \right\} \nonumber \\ 
 &= \mathrm{min}\left\{ 1, e^{(\beta_{m+1} - \beta_m) (\chi(\vv x_l^{m+1})-\chi(\vv x_l^{m}))} \right\} \,,
 \label{eq:pswap}
\end{align}
with the Boltzmann distribution for each $\beta_m$ given by $p^m(\vv{x}) = 1/Z_m\exp\left(-\chi(\vv{x})\beta_m\right)$. Such swap moves are conveniently proposed after a certain number of sweeps, which satisfies the sufficient condition of balance for thermodynamics~\cite{ea.de.05}. In practice, we chose 10 sweeps before swapping replicas. For the exchange to effectively take place the underlying requirement is that the adjacent $\beta_m$ and $\beta_{m+1}$ values are close enough to each other, so that the two energy distributions
 $\Omega[\chi(\vv x)]p^m(\vv{x})$ and  $\Omega[\chi(\vv x)]p^{m+1}(\vv{x})$ overlap, with 
\mbox{$\Omega[\chi_0]=\int d\vv{x}\ \delta(\chi_0-\chi(\vv{x})) $} 
the density of states of the cost function.
 This means that a replica at one temperature must represent a likely configuration for the neighboring temperature~\cite{ea.de.05,kofk.02}. In order to achieve this, a crucial point in the PT algorithm is to adjust the distribution of the inverse temperatures properly to the considered situation. Various criteria for this have been devised, see e.g. \tcite{ea.de.05}. A common choice is to demand that the swapping probability \eeqref{eq:pswap} becomes constant as a function of temperature~\cite{ra.ch.05,ko.ko.05}, and in \tcite{ka.tr.06} a feedback strategy was presented which optimizes the round trip times of replicas. We tested the latter within AMEA but favored the simpler former criterion in the end, since it allows for a rapid feedback and quick adjustment to large changes in $\chi(\vv x_l^{m})$. In the simple situation of a constant specific heat $C_H$ with respect to energy $\chi$ for instance, an optimal strategy is known since a geometric progression $\beta_m/\beta_{m+1} = \mathrm{const.}$ of temperatures yields a constant swapping probability~\cite{ea.de.05,kofk.02}. For interesting cases in practice this is rarely fulfilled, but within AMEA it served as a good starting point. The set of inverse temperatures is then optimized by averaging $p_{\mathrm{swap},l}^{m,m+1}$ over a couple of swappings to obtain the mean probability $\bar p_{\mathrm{swap}}^{m,m+1}$ and adjusting the $\beta_m$ thereafter. For this we chose a fixed lowest and highest $\beta_m$ value and changed the spacings in between according to 
\begin{equation}
 \Delta \beta_m' = c  \frac{\Delta \beta_m}{\log\left(\bar p_{\mathrm{swap}}^{m,m+1}\right)} \,, 
 \label{eq:optTK}
\end{equation}
with $\Delta \beta_m = \beta_{m+1} - \beta_m$ and $c$ adjusted properly so that $\max(\beta_m')-\min(\beta_m') = \max(\beta_m) - \min(\beta_m)$. In the works by \tcite{ra.ch.05,ko.ko.05} it was shown that a constant swapping probability of $20\%-23\%$ seems to be optimal. We determined the highest and lowest $\beta_m$ values by the changes in $\chi(\vv{x})$ we want to resolve or allow for, and the number of inverse temperatures $\beta_m$ was then set accordingly in order to roughly obtain $\bar p_{\mathrm{swap}}^{m,m+1} \approx 0.25$. Fixing the smallest and largest $\beta_m$ 
is, for our purposes, the most convenient choice among the many possibilities.

However, despite of the feedback optimization of temperatures as just described above, we encountered in practice the unwanted behavior that the set of parallel replicas effectively decoupled into several clusters. In order to suppress this we found it advantageous to introduce the following simple modification to \eeqref{eq:pswap}~\footnote{ 
One should note that the modification violates balance conditions and therefore the applicability in statistical physics. However, it is perfectly valid for the purpose of minimization problems.
}
\begin{equation}
 p_{\mathrm{swap},l}^{m,m+1}  = \max\left\{ p_{\mathrm{swap},l}^{m,m+1} , p_\mathrm{swap}^\mathrm{th.}\right\} \,,
 \label{eq:pswapnew}
\end{equation}
with a certain threshold probability $p_\mathrm{swap}^\mathrm{th.}$, e.g. $p_\mathrm{swap}^\mathrm{th.} = 0.1$ or $0.05$. In this way one avoids that the $\beta_m$ are shifted unnecessarily close to each other and avoids very long time scales, in which replicas oscillate only between two neighboring inverse temperatures.

\section{Matrix form and number of independent parameters for the different setups}
\label{mat}

\newcommand{\ppp}[1]{x_{#1}}
\newcommand{\mparagraph}[1]{\subsection*{#1}}

For the sake of clarity we present here for the different setups of 
Fig.~\ref{fig:setups_sketch}
the form of the (hermitian) matrices $\vv E$ and $\vgam1$ for the case $N_B=4$ in the particle-hole symmetric case, i.e. under the constraint \eqref{egph} which also fixes $\vgam2$.
 In addition, we quote the number of available fit parameters $C(N_B)$ for each setup.
The fit parameters are denoted below as
$\ppp{i}$ for $ i=1,C(N_B)$, 
 with the only constraint that $\vgam{i}$ should be semipositive definite.
This, together with 
the requirement that $\und \Delta_\aux$ vanishes for  $\omega\to\infty$ further requires $\Ga{1/2}_{ff}=\Ga{1/2}_{if}=\Ga{1/2}_{fi}=0$.
In the first four setups, the impurity is in the center ($i=3$). In the ``1 chain n.n.'' it is on the first site ($i=1$).

\mparagraph{``full'' geometry}

\beq
\label{et}
\vv E = \vv E_t \equiv
 \begin{pmatrix}
\ppp1 & \ppp3 & 0 & 0  & 0  \\
\ppp3 & \ppp2 & \ppp4 & 0  & 0  \\
0 & \ppp4 & 0 & \ppp4  & 0  \\
0 & 0 & \ppp4 & -\ppp2 & \ppp3  \\
0 & 0 & 0 & \ppp3  & -\ppp1
\end{pmatrix}
\eeq

\beqs
\vgam1 =
 \begin{pmatrix}
\ppp5  & \ppp9  & 0 & \ppp{11} & \ppp{12} \\
\ppp9^*  & \ppp6  & 0 & \ppp{13} & \ppp{14} \\
0  & 0  & 0 & 0  & 0  \\
\ppp{11}^* & \ppp{13}^* & 0 & \ppp7  & \ppp{10} \\
\ppp{12}^* & \ppp{14}^* & 0 & \ppp{10}^* & \ppp8 
\end{pmatrix}
\eeqs
The parameters  $\ppp9$ to $\ppp{14}$ can be complex.
Therefore, it is straightforward to see that, for general $N_B$
 the number of independent (real) parameters is
$
C(N_B)=\frac{N_B}{2}*(N_B+3) 
$
 for the real case 
and 
$
C(N_B)=N_B*(N_B+1) 
$
 for the complex case.

\mparagraph{``2-chain n.n.'' geometry}
\beqs
\vv E = \vv E_t  \quad \text{Eq.~\eqref{et}}
\eeqs

\beqs
\vgam1=
 \begin{pmatrix}
\ppp5  & \ppp9  & 0 & 0 & 0 \\
\ppp9  & \ppp6  & 0 & 0 & 0 \\
0  & 0  & 0 & 0  & 0  \\
0 & 0 & 0 & \ppp7  & \ppp{10} \\
0 & 0 & 0 & \ppp{10} & \ppp8 
\end{pmatrix}
\eeqs
so in general $C(N_B)=3N_B-2$ 

\mparagraph{``2-chain onsite'' geometry}
\beqs
\vv E = \vv E_t \quad  \text{Eq.~\eqref{et}}
\eeqs

\beqs
\vgam1 = 
 \begin{pmatrix}
\ppp5  & 0  & 0 & 0 & 0 \\
0 & \ppp6  & 0 & 0 & 0 \\
0  & 0  & 0 & 0  & 0  \\
0 & 0 & 0 & \ppp7  & 0 \\
0 & 0 & 0 & 0 & \ppp8 
\end{pmatrix}
\eeqs
Here, $C(N_B)=2N_B$  

\mparagraph{``star'' geometry}

\beqs
\vv E = \begin{pmatrix}
\ppp1 & 0 & \ppp3  & 0  & 0  \\
0 & \ppp2 & \ppp4  & 0  & 0  \\
\ppp3 & \ppp4 & 0  & \ppp4  & -\ppp3 \\
0 & 0 & \ppp4  & -\ppp2 & 0  \\
0 & 0 & -\ppp3 & 0  & -\ppp1
\end{pmatrix}
\eeqs

\beqs
 \vgam1 = \begin{pmatrix}
   \ppp5 & 0          & 0         & 0      & 0  \\
   0     & \ppp6      & 0         & 0      & 0  \\
    0    & 0          & 0         & 0      & 0 \\
0        & 0          & 0         & \ppp7  & 0  \\
0        & 0          & 0         & 0      & \ppp8
\end{pmatrix}
\eeqs
Also here $C(N_B)=2N_B$.

\mparagraph{``1 chain n.n.'' geometry}
Remember, here the impurity is on $i=1$.
\beqs
\vv E =
 \begin{pmatrix}
 0 & \ppp1 & 0 & 0 & 0 \\
\ppp1 & 0 & \ppp2 & 0 & 0 \\
0 & \ppp2 & 0 & \ppp3 & 0 \\
0 & 0 & \ppp3 & 0 & \ppp4 \\
0 & 0 & 0 & \ppp4 & 0
\end{pmatrix}
\eeqs

\beqs
\vgam1=
 \begin{pmatrix}
0  & 0  & 0 & 0 & 0 \\
0  & \ppp5  & \ppp9 & 0 & 0 \\
0  & \ppp9  & \ppp6 & \ppp{10}  & 0  \\
0 & 0 & \ppp{10} & \ppp7  & \ppp{11} \\
0 & 0 & 0 & \ppp{11} & \ppp8 
\end{pmatrix}
\eeqs
In this case, $C(N_B)=3N_B-1$. 

\section{Reduction of bath to a ``star'' form}
\newcommand{\wT}{\widehat{\vv T}}
\newcommand{\DRauxo}{\Delta^R_\mathrm{aux}}
\label{rstar}
In principle, one can represent a noninteracting dissipative bath consisting of
$N_B$ sites ($i=1,\dots N_B)$ coupled to an impurity ($i=f$, we take  $f=0$) by specifying
the single-particle parameters $E_{ij}$,  
$\Ga1_{ij}$, and $\Ga2_{ij}$ ($i,j=0,\dots N_B$), with corresponding hermitian, and in the case of $\vgam1$, $\vgam2$ semipositive definite matrices. 
We show here that for the sake of fitting 
the {\em retarded} component of 
a given bath spectral function
$\DRauxo$, these parameters
are redundant.

We rewrite \eeqref{gro}  in block form
\begin{equation}
\VG^R=\left(\begin{array}{cc}
\omega-F_{0}\:\: & - \wT\\
-\vv T & \omega-\vv F
\end{array}\right)^{-1}
\label{grf}
\end{equation}
where the first $1\times1$ block contains~\footnote{In order to be more general, we allow for 
nonzero elements of the $\vv \Gamma$ matrices on the impurity site as well}
$F_{0}\equiv E_{00} -i\Ga{+}_{00}$,
the $N_B\times N_B$ complex
matrix $\vv F$ is given by $F_{ij}\equiv E_{ij}-i\Ga{+}_{ij}$
for $i,j=1,\dots N_B$,  the column vectors 
$T_{i}\equiv E_{i0}-i\Ga{+}_{i0}$, 
$\widehat T_{i}\equiv E_{0i}-i\Ga{+}_{0i}$, 
and we have introduced \mbox{$\vgam{\pm}\equiv\vgam{1}\pm\vgam{2}$}.

We are interested in $G_\aux^R$, which is the ${}_{00}$ component of $\VG^R$.
By a well known result of matrix inversion, this is given by 
\begin{equation}
1/G_\aux^R=\left(\omega-F_{0}-\wT(\omega-\vv F)^{-1}\vv T\right)\label{1gr}\;,
\end{equation}
 which identifies $\DRauxo=\vv T^{T}(\omega-\vv F)^{-1}\vv T+\delta F_{0}$,
where $\delta F_{0}\equiv F_{0}- \eps_f$, which, for simplicity, we set to zero. The first term can be rewritten
by introducing the matrix $\vv V$ which diagonalizes $\vv F$,~\footnote{Note that diagonalisation of $\vv F$ is not always guaranteed}
 i.e. 
\beq
\label{uf}
\vv V^{-1}\vv F \vv V=\vv F_{diag}\;.
\eeq
This gives
\beqa
\nonumber 
\DRauxo& =\wT\vv V {\vv V}^{-1}(\omega-\vv F)^{-1}\vv V {\vv V}^{-1}\vv T
\\ & \nonumber 
=\bbar{\wT} (\omega-{\vv F}_{diag})^{-1}\bbar{\vv T} 
\\  \nonumber 
\bbar{\vv T} &\equiv {\vv V}^{-1} \vv T 
\quad \quad
\bbar{\wT} \equiv \wT {\vv V}
\;.
\eeqa
We can thus replace in \eeqref{grf} $\vv F$ with a diagonal, complex matrix $\vv F_{diag}$
and $\vv T$ ($\wT$)
with  $\bbar{\vv T}$
($\bbar{\wT}$), and we get
\beqa
\VG^{\prime R} &= \left(\omega - \vv F'\right)^{-1} \nonumber \\
\vv F' &\equiv 
\left(\begin{array}{cc}
F_{0}\:\: &  {\bbar{\wT}}\\
{\bbar{\vv T}} & \vv F_{diag}
\end{array}\right)^{-1} \;.
\eeqa
Here, $\VG^{\prime R}$
 has the same ${}_{00}$ element as $\VG^R$ from \eeqref{grf}, i.e. the same $G^R_\aux$ and $\DRauxo$.
In this way, by the requirement that $\vv E$ and $\vgam{+}$ must be hermitian, we can construct new 
$\vv E' = (\vv F' +  \vv F^{\prime\dag})/2$ and
${\vv\Gamma}^{\prime(+)} = (\vv F' -  \vv F^{\prime\dag})/(2i)$, i.e. a new auxiliary system 
 which {\em  yield the same $\DRauxo$} and
have the ``star'' geometry (cf. \ref{fig:setups_sketch}).~\footnote{Interestingly, in many cases the new ${\vv\Gamma}^{\prime(+)}$ is no longer positive definite. 
}
This means that, concerning the {\em retarded part,} one
can restrict to the case of diagonal bath energies and $\vgam{+}$, i.e., as in the non-dissipative case,
$\DRauxo$ is fixed by
 only 
$\mathcal{O}(N_B)$ independent bath parameters, the rest being redundant.
This is also the case when the bath hybridisation function is represented by a completely empty and a completely full chain, as discussed in Sec.~\ref{sec:results}, since in that case one simply fits the {\em retarded} components of the two chains separately. 
On the other hand, 
for the most generic case, 
  $\vgam{1}$ and $\vgam{2}$  will not commute and cannot be simultaneously diagonalized, so that 
the Keldysh component $\DKaux$ appears to still depend on 
$\mathcal{O}(N_B^2)$ bath parameters (cf. Fig.~\ref{fig:conv_chi_NB}). 
Further investigations should be carried out in order to clarify this issue. 
\bibliography{references_database}
\bibliographystyle{prsty}

\end{document}